\documentclass{sig-alternate}
\usepackage{mathptmx} 

\usepackage{fancyhdr}
\usepackage[normalem]{ulem}
\usepackage[hyphens]{url}
\usepackage[sort,nocompress]{cite}
\usepackage[final]{microtype}
\usepackage[keeplastbox]{flushend}
\usepackage[bookmarks=true,breaklinks=true,letterpaper=true,colorlinks,linkcolor=black,citecolor=blue,urlcolor=black]{hyperref}
\usepackage{url}
\usepackage{threeparttable}
\usepackage{tikz}
\usepackage{amsmath}
\usepackage{booktabs}
\usepackage{subfig}
\usepackage{comment}
\usepackage{pifont}
\usepackage{soul}
\usepackage{multirow}

\newcommand{\squishlist}{
   \begin{list}{$\bullet$}
    { 
    \setlength{\itemsep}{0pt}      \setlength{\parsep}{0pt}
      \setlength{\topsep}{3pt}       \setlength{\partopsep}{0pt}
      \setlength{\listparindent}{-2pt}
      \setlength{\itemindent}{-5pt}
      \setlength{\leftmargin}{1em} \setlength{\labelwidth}{0em}
      \setlength{\labelsep}{0.5em} } }

\newcommand{\squishend}{
    \end{list}  }

\fancypagestyle{firstpage}{
  \fancyhf{}
  
  \fancyhead[C]{\vspace{15pt}\normalsize{}} 
  \fancyfoot[C]{\thepage}
}

\title{Demystifying Arch-hints for Model Extraction: An Attack in Unified Memory System} 
\begin{document}

\author{%
\and
Zhendong Wang\\
  \affaddr{University of Texas at Dallas}\\
  \affaddr{Richardson, Texas, USA}\\
  \email{Zhendong.Wang@utdallas.edu}
\and
Xiaoming Zeng\\
  \affaddr{University of Texas at Dallas}\\
  \affaddr{Richardson, Texas, USA}\\
  \email{Xiaoming.Zeng@utdallas.edu}
\and
Xulong Tang\\
  \affaddr{University of Pittsburgh}\\
  \affaddr{Pittsburgh, Pennsylvania, USA}\\
  \email{tax6@pitt.edu}
\and
Danfeng Zhang\\
  \affaddr{The Pennsylvania State University}\\
  \affaddr{University Park, Pennsylvania, USA}\\
  \email{zhang@cse.psu.edu}
\and
Xing Hu\\
  \affaddr{Institute of Computing Technology}\\
  \affaddr{Beijing, China}\\
  \email{huxing@ict.ac.cn}
\and
Yang Hu\\
  \affaddr{Tsinghua University}\\
  \affaddr{Beijing, China}\\
  \email{hu\_yang@tsinghua.edu.cn}
}

\maketitle
\thispagestyle{firstpage}
\pagestyle{plain}


\begin{abstract}
\begin{sloppypar}
The deep neural network (DNN) models are deemed confidential due to their unique value in expensive training efforts, privacy-sensitive training data, and proprietary network characteristics.   
Consequently, the model value raises incentive for adversary to steal the model for profits, such as the representative model extraction attack. 
Emerging attack can leverage timing-sensitive architecture-level events (i.e., Arch-hints) disclosed in hardware platforms to extract DNN model layer information accurately. 
In this paper, we take the first step to uncover the root cause of such Arch-hints and summarize the principles to identify them. We then apply these principles to emerging Unified Memory (UM) management system and identify three new Arch-hints caused by UM's unique data movement patterns. We then develop a new extraction attack, UMProbe. We also create the first DNN benchmark suite in UM and utilize the benchmark suite to evaluate UMProbe. 
Our evaluation shows that UMProbe can extract the layer sequence with an accuracy of 95\% for almost all victim test models, which thus calls for more attention to the DNN security in UM system.

\end{sloppypar}









\end{abstract}
\vspace{-3pt}
\section{Introduction}
\vspace{-3pt}
The ever-increasing employment of machine learning (ML) technology, especially deep neural networks (DNNs), in various applications has tremendously improved these application domains' efficiency, such as computer vision \cite{CV1,he2016deep,simonyan2014very}, speech recognition \cite{speech1,speech3}, natural language processing \cite{nlp1,nlp2}, and etc.
These ML and DNN models possess unique value, which is reflected in the expensive training efforts, privacy-sensitive training data, and proprietary network architectures and parameters. Thus, these models are usually deemed confidential as protected IPs.
Consequently, these confidential models make them appealing targets to the adversary who intends to steal the models for profits \cite{chen2020stealing}. 
The adversary can explore the model execution and infer the non-disclosed model architecture and parameters through extraction attacks. Such attack is known as \textbf{\textit{model extraction attacks}}\cite{batina2019csi, duddu2018stealing, hu2020deepsniffer, hua2018reverse, naghibijouybari2018rendered, yan2020cache}, which not only destroys the confidentiality of a model and damages the IP of the owner but also benefits further attacks \cite{liu2016delving, oh2019towards}.

ML and DNN models are mainly deployed in cloud with publicly accessible query interfaces/APIs, known as ML-as-a-service (MLaaS), allowing users to obtain service (e.g., predictive analytics) without accessing the black-box models.
Thus, the adversary can duplicate the model functionality by exploring such attack surfaces as querying APIs \cite{tramer2016stealing}. 
Recently, with the proliferation of ML techniques in edge devices, such as autonomous driving \cite{av1,av2}, model extraction attack has such effective approaches as hardware side-channel attacks (SCAs) to  pry into the model’s internal architecture. 
The prevalence of edge-deployed ML models and the pursuit of the extraction accuracy drive the adversaries to explore new attack surfaces instead of the superficial querying mode.

Prior works \cite{hua2018reverse, naghibijouybari2018rendered, hu2020deepsniffer} demonstrate the SCAs can capture certain architectural events or hardware behaviors (e.g., bus traffic through bus snooping) during model execution. These timing-sensitive architectural events or hardware behaviors can be leveraged by the adversary to infer the DNN layer architectures and perform accurate DNN model extraction attacks. We argue that such \textit{effective} architectural events, dubbed \textbf{\textit{Architecture hints (Arch-hints)}} in this paper, present a new \textbf{\textit{Hardware/Architecture-level attack surface}} for the model extraction in the edge/local deployment. Though existing work shed some light in utilizing architecture-level events and behaviors in GPU-based model extraction\cite{naghibijouybari2018rendered,wei2020leaky,hu2020deepsniffer}, these events are used in an ad-hoc manner. It still lacks a systematic exploration and formal definition of such architectural behaviors, which conceals the universality of such threat on different platforms.  

In this paper, we set out to investigate prior identified Arch-hints, uncover the root cause of Arch-hints, and clearly define the Arch-hints. The key insight is that we identify that these Arch-hints essentially result from \textit{\textbf{data movement in hardware platforms during model execution}}. 
Nevertheless, simply caused by tractable data movement during model execution doesn't entitle an architectural event to an Arch-hint. The data movement during model execution must also exhibit distinguishable and stable patterns across the DNN model layers to make it a qualified Arch-hint. 
 

Nowadays, the Graphics Processing Unit (GPU) has become the dominant hardware to deploy DNN applications, both in cloud and edge scenarios \cite{dnncloud2,dnncloud3,dnncloud4, li2018learning, yazici2018edge}. Also, the considerable memory footprint of DNN-based workloads and ever-increasing requirements of programming flexibility has pushed the GPU memory management on the verge of a major shift from the traditional copy-then-execute (CoE) model to the unified memory (UM) model \cite{li2015evaluation, landaverde2014investigation, otterness2017evaluation, umbeginners, bateni2020co, wang2020enabling, 2020hotedge}, especially on these memory-limited edge platforms \cite{bateni2020co, wang2020enabling, dashti2017analyzing}. 
Based on the principles to define Arch-hints, we identify three Primary Arch-hints that are specifically caused by the data movement patterns of UM, namely  \textit{page fault latency (PFLat), page migration latency (MigLat)}, and \textit{page migration size (MigSize)}, which exhibits distinguishable patterns on layer features and model architecture during model execution (Sec. \ref{UMbrief}).

We propose a metric \textit{effectiveness\_score} to validate the effectiveness of these Primary Arch-hints (Sec. \ref{UM-archhints}) by evaluating their distributions across the DNN model layers. 
Then, we propose a new model extraction attack, \textbf{UMProbe}, which thoroughly explores the new Arch-hints in UM system to perform model extraction accurately (Sec. \ref{newatk}). We also evaluate how existing Arch-hints and their combinations with Primary Arch-hints can affect the model extraction accuracy.

Lastly, we substantially modify the Darknet framework and develop the UM implementation for a portfolio of representative DNN benchmarks. To the best of our knowledge, no UM implementations of DNN models has been published. We evaluate UMProbe performance on these benchmarks and demonstrate that UMProbe can extract victim model layer sequence with the accuracy of 95\% for almost all victim models (Sec. \ref{evaluation}).
In summary, this paper makes following contributions:  
\squishlist{}
    \item We investigate prior identified Arch-hints, uncover the root cause of these Arch-hints, and formally define the Arch-hints. Based on the definition, we identify three primary Arch-hints cause by the unique data movement patterns of GPU UM system.
    \item We characterize multiple Arch-hints candidates in UM system and propose a metric to quantify their effectiveness.   
    \item The newly explored Arch-hints expose a new attack surface in UM which has not been explored before. We develop an extraction attack UMProbe based on that.
    \item We create the first DNN benchmark suite under UM to facilitate the related researches in the community.
    \item We evaluate UMProbe performance using the benchmark suite and demonstrate UMProbe's high accuracy, calling for attention to the DNN security in UM system.   
\squishend{}

\vspace{-3pt}
\section{Extracting Model Using Hardware Architectural Hints}
\vspace{-3pt}

\subsection{Model Extraction Essential}\label{attback}

Model extraction attacks target the ML models deployed in the cloud with publicly accessible query interfaces/APIs, which known as ML-as-a-Service (MLaaS) since its inception. The adversary tries to duplicate a functionally equivalent model by frequently querying APIs for cloud-based models. Recently, model extraction attack also extends to the ML models served on the edge and local devices with the proliferation of ML techniques in modern applications such as autonomous driving\cite{mlavs1,yang2019re,mlavs3,mlavs4,li2018learning, yazici2018edge, verhelst2020machine}. In this paper, we focus on this emerging trend and set out to explore the model extraction attack targeting the ML deployment in edge scenarios, which present higher threats to the models.


\noindent\textbf{Attack Target: What to Steal?}
As a DNN model consists of network architecture, parameters and hyper-parameters, the adversary can target extracting architecture \cite{oh2019towards, hu2020deepsniffer}, parameters \cite{tramer2016stealing}, hyper-parameters \cite{wang2018stealing}.
Specifically, architecture indicates layer types and connections. Parameters are weights and biases that are learned during training process. Hyper-parameters are the configuration variables used to control the training process, such as learning rate, batch size, etc.

Among all these targets, the \textit{network architecture is the most fundamental and valuable targets for a DNN model extraction} as both model parameters and hyper-parameters can be inferred with the knowledge of the model architecture \cite{tramer2016stealing, wang2018stealing}. 
The adversary can even launch adversarial attack based on the extracted network architecture \cite{liu2016delving, hu2020deepsniffer}.
The desired network architecture usually consists of layer number, layer types/dimension, and layer connections. The layer connection can include the sequential (e.g., VGG \cite{simonyan2014very}) and the non-sequential (e.g., shortcut in ResNet \cite{he2016deep}).

\vspace{-3pt}
\subsection{Attack Surface: How to Steal?}\label{attsurface}
\vspace{-3pt}
\noindent\textbf{Application/API-Level Attack Surface:}
Conventionally, the adversary performs extraction attack at \ul{\textit{application/API-level}}. In this attack surface, the adversary accesses the API by querying the victim model and receives the replies. It then leverages the input-output pairs of the victim model to detect the decision boundary (i.e., the classification boundary between different classes) of the model and subsequently duplicates the model \cite{oh2019towards, tramer2016stealing}. 
However, such attack needs tons of queries and consumes significant computation resources and time \cite{oh2019towards}. 
Moreover, the attack can only duplicate the functionality of the model instead of being able to probe the accurate internal architecture of the model \cite{oh2019towards, jagielski2020high} due to its limited access to the cloud-deployed black-box model, which can hardly satisfy adversary's appetite. Thus, new attack surface revealing accurate information on model internal architecture besides model functionality is needed.

\begin{figure}[t]
\centering
            \includegraphics[width=1\linewidth]{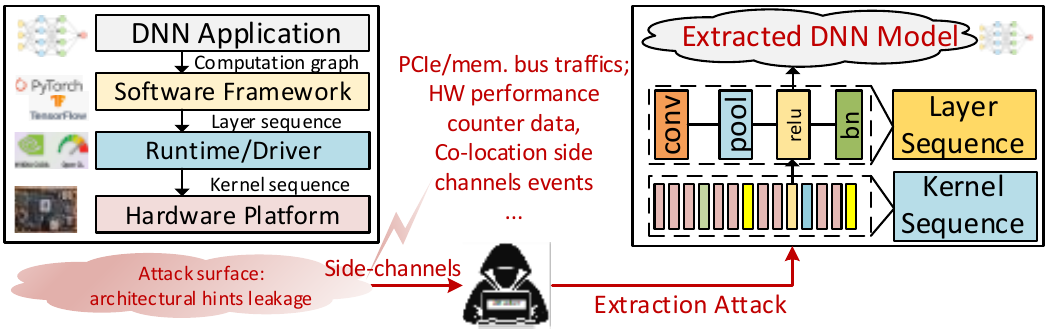}
            \vspace{-6mm}
            \caption{Extracting DNN models by exploring the attack surface provided by Architectural-hints.}\label{hintsandattack}
            \vspace{-4mm}
\end{figure}

\noindent\textbf{Hardware/Architecture-Level Attack Surface:}
The pursuit of the extraction accuracy and the prevalence of edge-deployed ML models drive the adversaries to explore new attack surfaces instead of the superficial querying mode \cite{hong2018security, hong20200wn, yan2020cache}. The hardware side-channel attacks (SCAs) have recently drawn attention since they can provide effective approaches to pry into the model's internal architecture. For example, \cite{hua2018reverse, naghibijouybari2018rendered, hu2020deepsniffer} demonstrate that SCAs can obtain information that is closely correlated to the model internal architecture by capturing certain architectural events and hardware behaviors during model execution. With further data analysis, the \ul{internal model architecture could be accurately inferred based on these critical hardware architectural behaviors}. We observe that such \ul{\textit{ hardware/architecture-level behavior leakage}} provides a new attack surface for model extraction attack. We name this hardware/architecture-level visible information as {\textit{Architecture-hints (Arch-hints)}}.



\subsection{Arch-hints for Model Extraction}\label{archhints}
In this work, we take the first step to present an in-depth exploration of how Arch-hints can contribute to the extraction attacks of the edge-deployed DNN models. Specifically, we illustrate a threat model with a GPU-based DNN inference setup. We investigate prior identified Arch-hints, analyze the root cause of these Arch-hints, and define the Arch-hints. Then, we use this definition to identify two critical Arch-hints in existing unified memory management systems, which can lead to new model extraction attacks towards edge-deployed DNN models. 

Fig.\ref{hintsandattack} shows an abstract view of how DNN model information translates to Arch-hints during the model execution. When the DNN application is executed in the DL framework (e.g., Pytorch), the framework formalizes the DNN model into a framework-level computational graph and then transforms the computational graph into the runtime layer \textit{execution sequence}, which is then issued to the runtime/hardware driver (e.g., CUDA, GPU driver). The runtime/hardware driver will launch a series of operational \textit{kernel sequence} accordingly. 
These kernel sequences could be revealed by carefully chosen Arch-hints while executing in the hardware platform (e.g., CPU-GPU heterogeneous platforms).

The adversary typically has physical access to the victim platforms and can co-locate its spy application in the same platforms. Thus, the adversary can capture these Arch-hints by leveraging hardware SCAs.
Prior works leverage architectural behaviors in model extraction attacks based on different platforms. Though these architectural behaviors share the similar functionalities as Arch-hints, they are used in an ad-hoc manner. It still lacks a systematic exploration and formal definition of such architectural behaviors. 

We summarize these Arch-hints and explore the hidden principles behind them. We categorize these Arch-hints into three types: a) cache-based, b) DRAM-based, c) GPU kernel-based, as shown in Table \ref{tb1:AvaiArchhints}. 
We illustrate the Arch-hints caught in GPU platforms. For instance,  \cite{naghibijouybari2018rendered} collects the GPU memory write transactions and GPU unified cache throughput as the Arch-hints. \cite{wei2020leaky} utilizes the Arch-hints including the number of GPU DRAM read/write requests and the number of GPU texture cache requests. \cite{hu2020deepsniffer} leverages Arch-hints such as memory bus traffics and kernel execution latency. 

\newcommand{\tabincell}[2]{\begin{tabular}{@{}#1@{}}#2\end{tabular}}
\begin{table}[t]
\scriptsize
	\centering \setlength\tabcolsep{1pt}
	\caption{Commonly-used Arch-hints in prior works.}
	\vspace{-1pt}
	\begin{tabular}{c|c|c|c|c|c|c}
		\hline
		\multirow{2}{*}{Attack}    & \multicolumn{3}{c|}{Arch-hints Used}  & \multirow{2}{*}{\tabincell{c}{Platform \& \\ Mem. Model}} &\multirow{2}{*}{Scenario} &\multirow{2}{*}{Approach} \\ \cline{2-4} 
		& Cache        & Memory   & Kernel  &\multicolumn{1}{c|}{} &\\ \bottomrule
		\tabincell{c}{RenderedInsecure \cite{naghibijouybari2018rendered}}   &\tabincell{l}{\checkmark} &\tabincell{l}{\checkmark} &\tabincell{l}{}& \tabincell{l}{GPU, CoE} & \tabincell{l}{Cloud/Edge} & \tabincell{l}{Predict}\\\hline
		\tabincell{c}{LeakyDNN \cite{wei2020leaky}}   &\tabincell{l}{\checkmark} &\tabincell{l}{\checkmark} &\tabincell{l}{}& \tabincell{l}{GPU, CoE} & \tabincell{l}{Cloud} & \tabincell{l}{Predict}\\\hline
		\tabincell{c}{DeepSniffer \cite{hu2020deepsniffer}}   &\tabincell{l}{} &\tabincell{l}{\checkmark} &\tabincell{l}{\checkmark}& \tabincell{l}{GPU, CoE} & \tabincell{l}{Edge} & \tabincell{l}{Predict} \\\hline
		\tabincell{c}{DeepRecon \cite{hong2018security}}   &\tabincell{l}{\checkmark} &\tabincell{l}{} &\tabincell{l}{}& \tabincell{l}{CPU, N/A} & \tabincell{l}{Cloud} & \tabincell{l}{Predict} \\\hline
		\tabincell{c}{0wnNAS \cite{hong20200wn}}   &\tabincell{l}{\checkmark} &\tabincell{l}{} &\tabincell{l}{}& \tabincell{l}{CPU, N/A} & \tabincell{l}{Cloud} & \tabincell{l}{Infer}\\\hline
		\tabincell{c}{StealNN \cite{duddu2018stealing}}   &\tabincell{l}{} &\tabincell{l}{} &\tabincell{l}{\checkmark}& \tabincell{l}{CPU, N/A} &\tabincell{l}{Cloud} &  \tabincell{l}{Predict} \\\hline
		\tabincell{c}{Cachetelepathy \cite{yan2020cache}}   &\tabincell{l}{\checkmark} &\tabincell{l}{} &\tabincell{l}{}& \tabincell{l}{CPU, N/A} & \tabincell{l}{Cloud} & \tabincell{l}{Search}\\\hline
		\tabincell{c}{ReverseCNN \cite{hua2018reverse}}   &\tabincell{l}{} &\tabincell{l}{\checkmark} &\tabincell{l}{\checkmark}& \tabincell{l}{FPGA, N/A} & \tabincell{l}{Cloud} & \tabincell{l}{Search}\\\bottomrule
	\end{tabular}
	\vspace{-5mm}
	\label{tb1:AvaiArchhints}
\end{table}

We delve into these Arch-hints and observe that these Arch-hints are essentially resulted from the \textit{data movement that exhibits in the hardware platforms} during model execution. For example, it is the data movement between GPU memory and GPU cache that causes the Arch-hint of memory bus traffic. The data movement between the GPU memory hierarchy system and GPU SMs that significantly contributes to the kernel execution latency.

While data movement is crucial to model execution exhibiting Arch-hints, we further identify that not all architectural behaviors caused by data movement can serve as effective "Arch-hints" of being leveraged in the attack.  
In fact, 
valid Arch-hints should be architectural events and hardware behaviors that present certain recognizable information for the adversary.

Based on the analysis above, Arch-hints is defined as effective architectural events and hardware behaviors that{\textit{1) being caused by tractable data movement during model execution, and 2) being able to exhibit recognizable information in extraction attack.}} 
We will utilize the definition to explore new Arch-hints in GPU unified memory system. 
\vspace{-3pt}
\section{Demystifying Arch-hints in Unified Memory}\label{archhintcharacter}
\vspace{-3pt}
Unified memory (UM) has gained widely adoption today due to its efficient memory footprint and programmability. In this section, we identify two unique Arch-hints in UM management system based on the definition proposed in Sec.\ref{archhints} and validate their effectiveness.

\vspace{-3pt}
\subsection{GPU Execution and Unified Memory} \label{UMbrief}
\vspace{-3pt}
We first introduce background of GPU execution model and unified memory management. As a representative vendor, NVIDIA GPU consists of several Streaming Multiprocessors (SMs), on-chip L2 cache, and high-bandwidth DRAM. 
All SMs share the unified L2 cache and the device memory through an on-chip interconnection network. In a typical discrete GPU setup, the GPUs are connected to the host CPU through PCIe interconnect. Note that, the discrete GPU has its own on-board physical memory which is physically separated from the CPU host memory. Since the GPU memory usually has less capacity compared to the CPU memory (e.g., 32 GB in NVIDIA V100~\cite{v100} v.s. hundreds GB of host CPU memory), the conventional GPU workload execution pattern is to copy the data from CPU memory to the GPU memory when needed, and copy the data back to CPU memory after the computation finishes.  
This execution model is referred to as \ul{\textit{copy-then-execute (CoE)}}. 

\begin{figure}[t]
\centering
            \includegraphics[width=\linewidth]{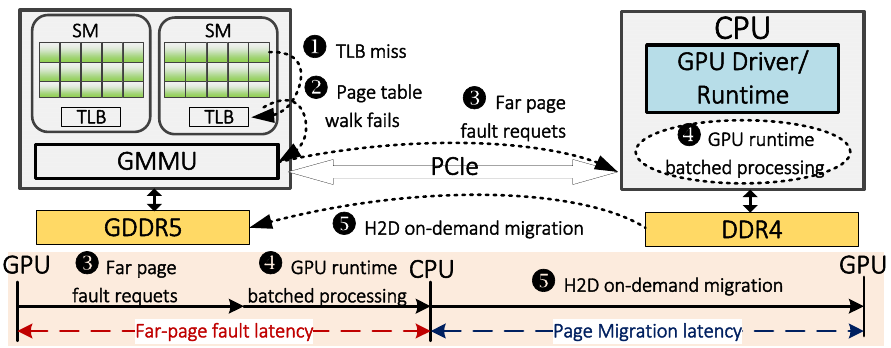}
            \caption{Far page fault and page migration in UM system.}
            \vspace{-4mm}
            \label{umprin}
\end{figure}

\noindent\textbf{Unified Memory Model:}
However, CoE execution mechanism suffers from i) frequent data copy between CPU and GPU~\cite{bateni2020co} and ii) out-of-memory problem due to limited GPU memory capacity~\cite{ganguly2019interplay}.
To address these disadvantages, \ul{\textit{ Unified Memory (UM)}} model is introduced to ease GPU programming by removing the necessity of explicit data copy between the CPU and GPU \cite{umbeginners,umamd}. Specifically, UM provides the illusion of unified virtual memory space for both CPU and GPU, and allows applications to access data on both CPU memory and GPU memory through a single shared pointer in the program. In CUDA programming, the API $cudaMallocManaged()$ is used to allocate UM space. Unlike CoE model which transfers the data by chunks, UM employs an on-demand paging method and transfers/migrates the data at page-level granularity. Fig.\ref{umprin} shows the data processing flow under UM model.
At system level, the GPU's page table walk will fail if SMs try to access a physical memory page that is not currently available in GPU local memory (i.e., the address translation lookup misses in both TLB and page table, steps \ding{172} $\sim$ \ding{173}) and a \ul{\textit{ far page fault}} exception 
is raised (step \ding{174}) by the GPU memory management unit (MMU)~\cite{zheng2016towards}. These exceptions are sent to the host CPU and handled by the driver (step \ding{175}). In particular, the driver first interrupts the CPU to handle the page faults, and then {\it migrates} the requested pages to the GPU memory (step \ding{176}) \cite{ganguly2019interplay}. We claim the system with CoE management model as CoE system. Accordingly, UM system indicates the system with UM management model.

\subsection{Arch-hints for UM System} \label{UM-archhints}

\begin{table*}[t]
\centering
\small
\caption{The collected and characterized candidate Arch-hints during DNN execution in UM system.}
\vspace{-10pt}
\begin{tabular}{c|c|c|c}
\hline
\tabincell{l}{Platform} &\tabincell{l}{Memory Hierarchy} &\tabincell{l}{Network Model} &\tabincell{l}{Collected candidate Arch-hints (8)}\\\hline
\tabincell{c}{Titan RTX \\(GPU)} &\tabincell{c}{Unified Memory \\(UM)}&\tabincell{c}{Darknet \\Reference \cite{darknetclassification}}&\tabincell{l}{L2 write transaction, L2 read transaction, DRAM read transaction, DRAM write transaction, \\ kernel latency, far fault latency, data migration latency, migration size}\\\hline
\end{tabular}
\vspace{-10pt}
\label{collectarchhints}
\end{table*}

\begin{figure*}[t]
\subfloat[L2 write trans. (Byte).]{
            \includegraphics[width=4.2cm, height = 3.2cm]{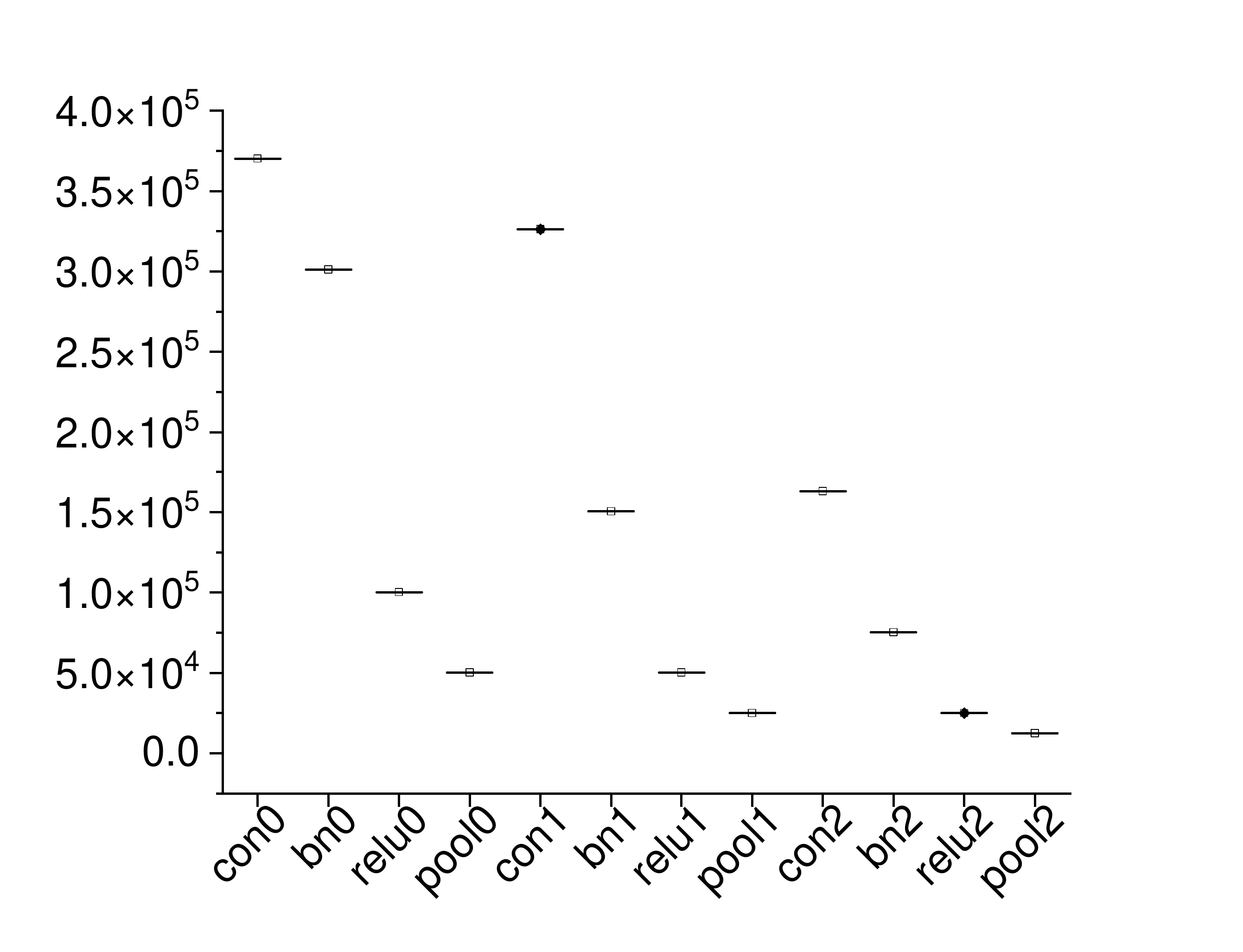} 
            \label{l2write}
}
\subfloat[DRAM write trans. (Byte).]{
            \includegraphics[width=4.2cm, height = 3.2cm]{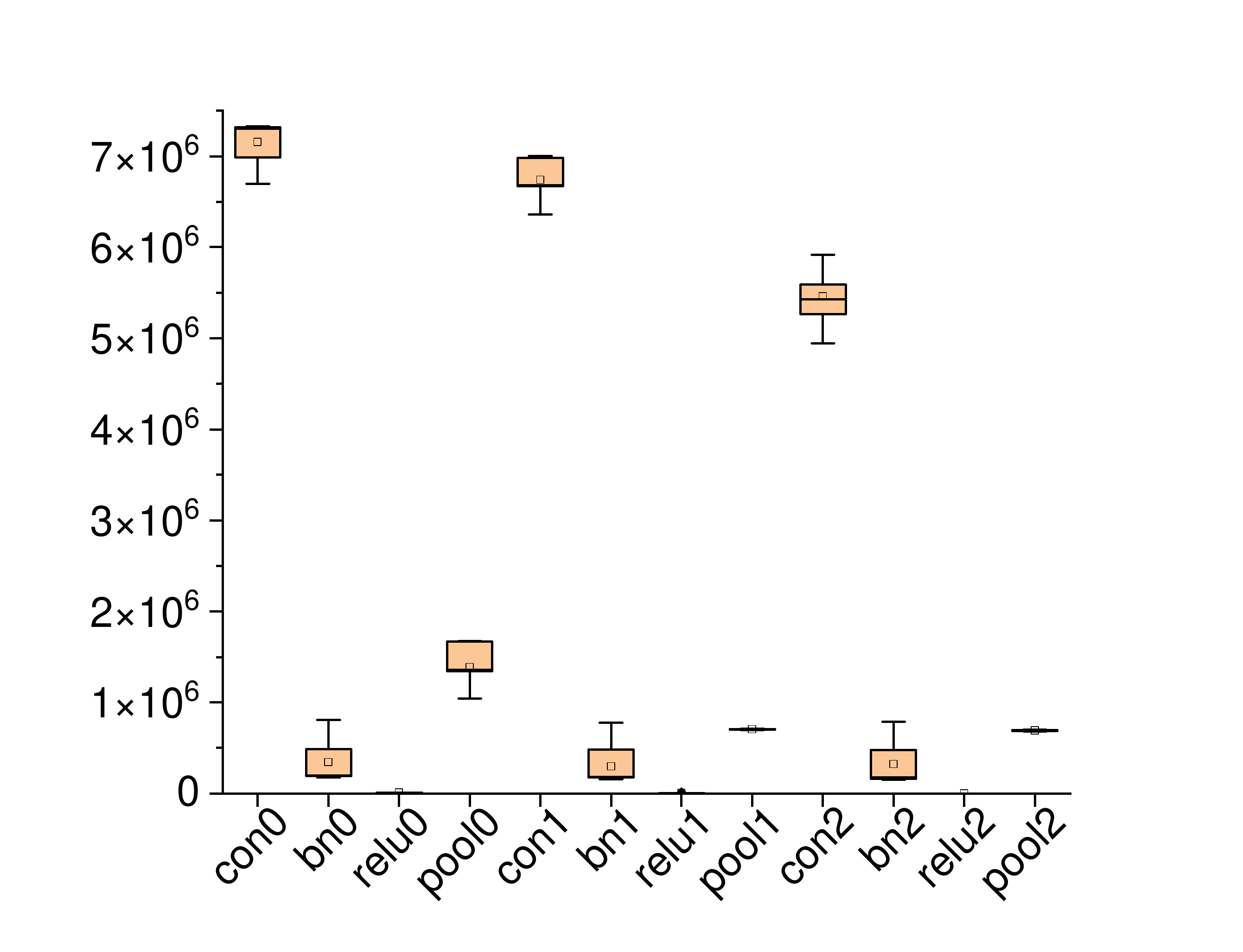} 
            \label{dramwrite}
}
\subfloat[L2 read trans. (Byte).]{
            \includegraphics[width=4.2cm, height = 3.2cm]{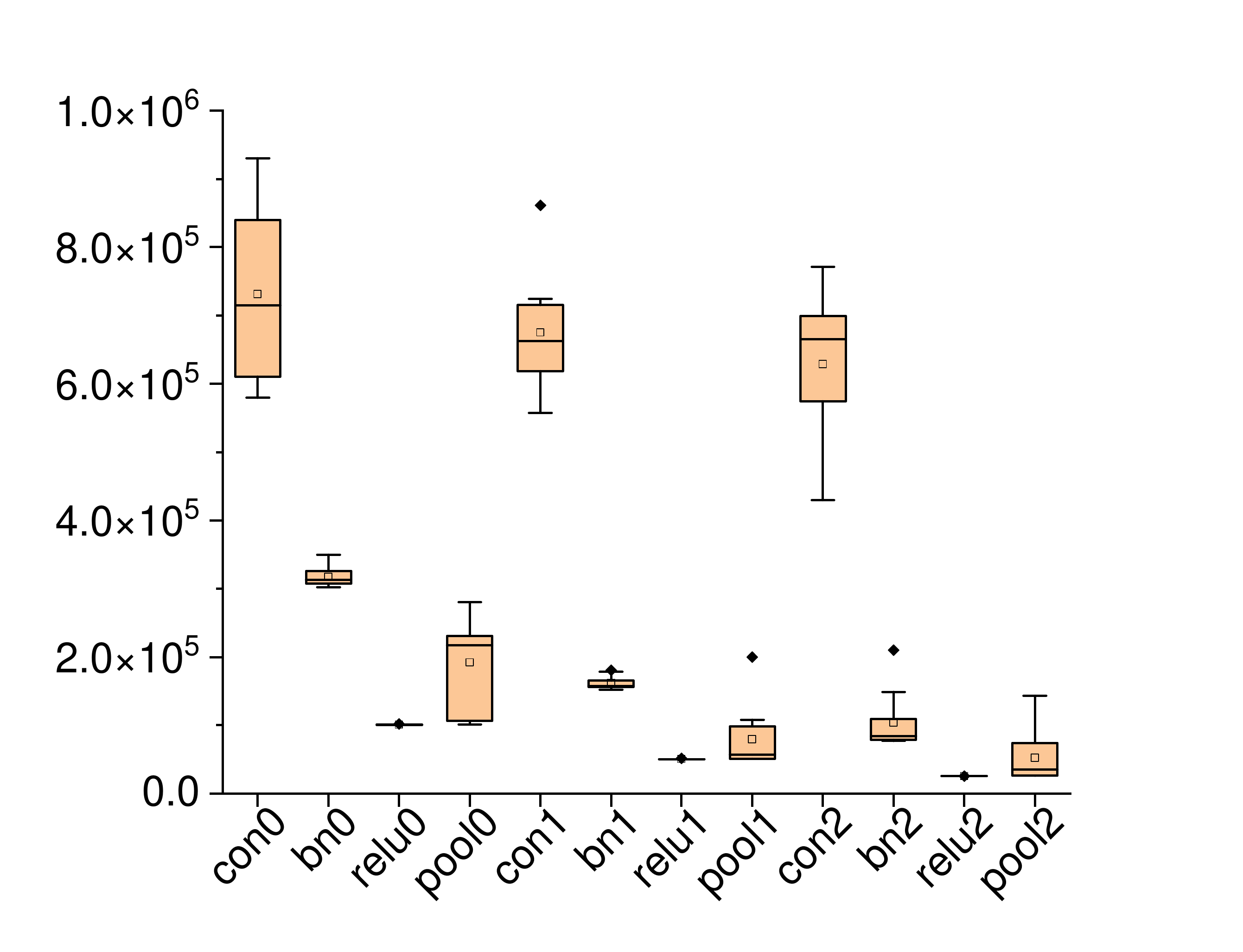} 
            \label{l2read}
}
\subfloat[DRAM read trans. (Byte).]{
            \includegraphics[width=4.2cm, height = 3.2cm]{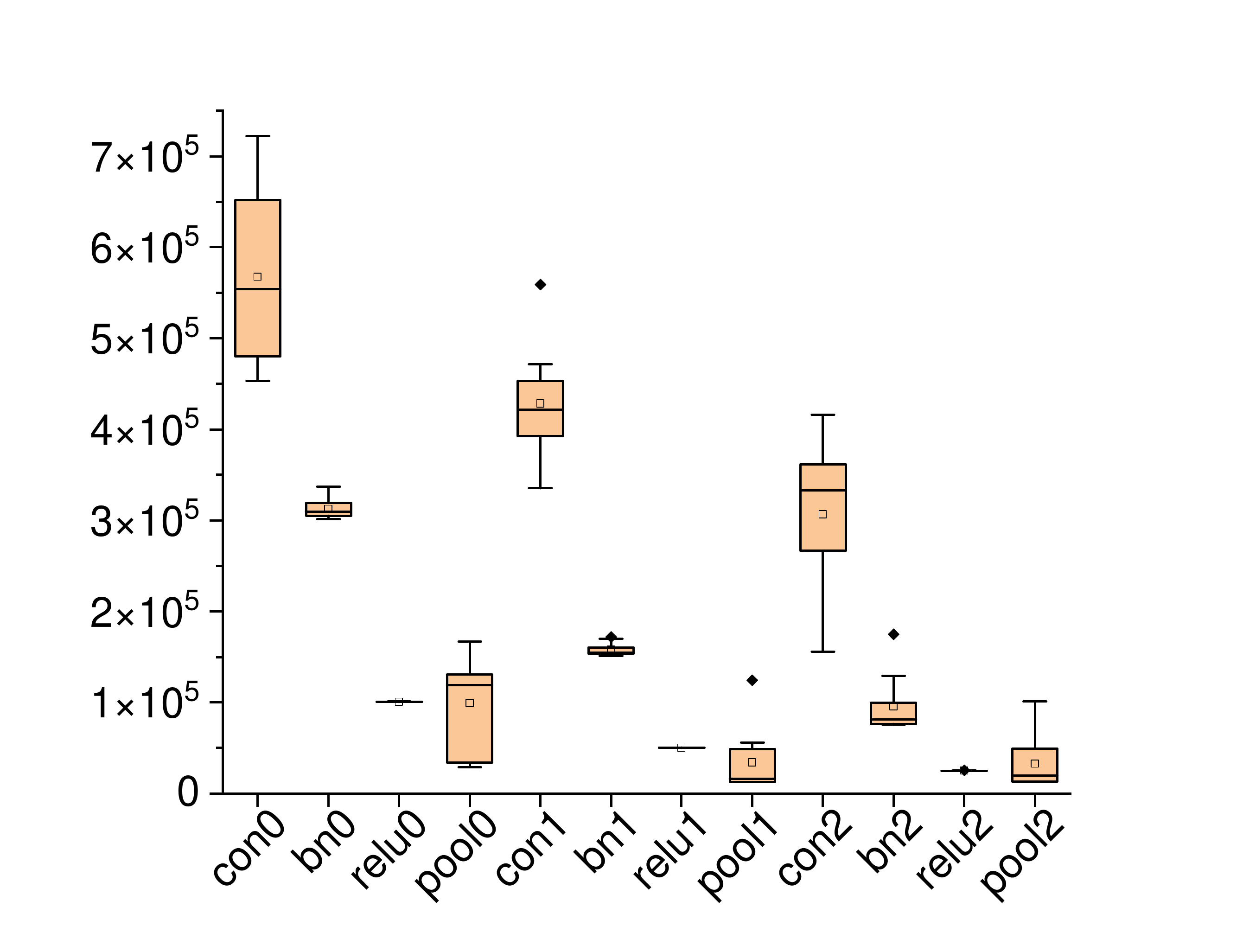} 
            \label{dramread}
}

\subfloat[Kernel latency (ms).]{
          \includegraphics[width=4.2cm, height = 3.2cm]{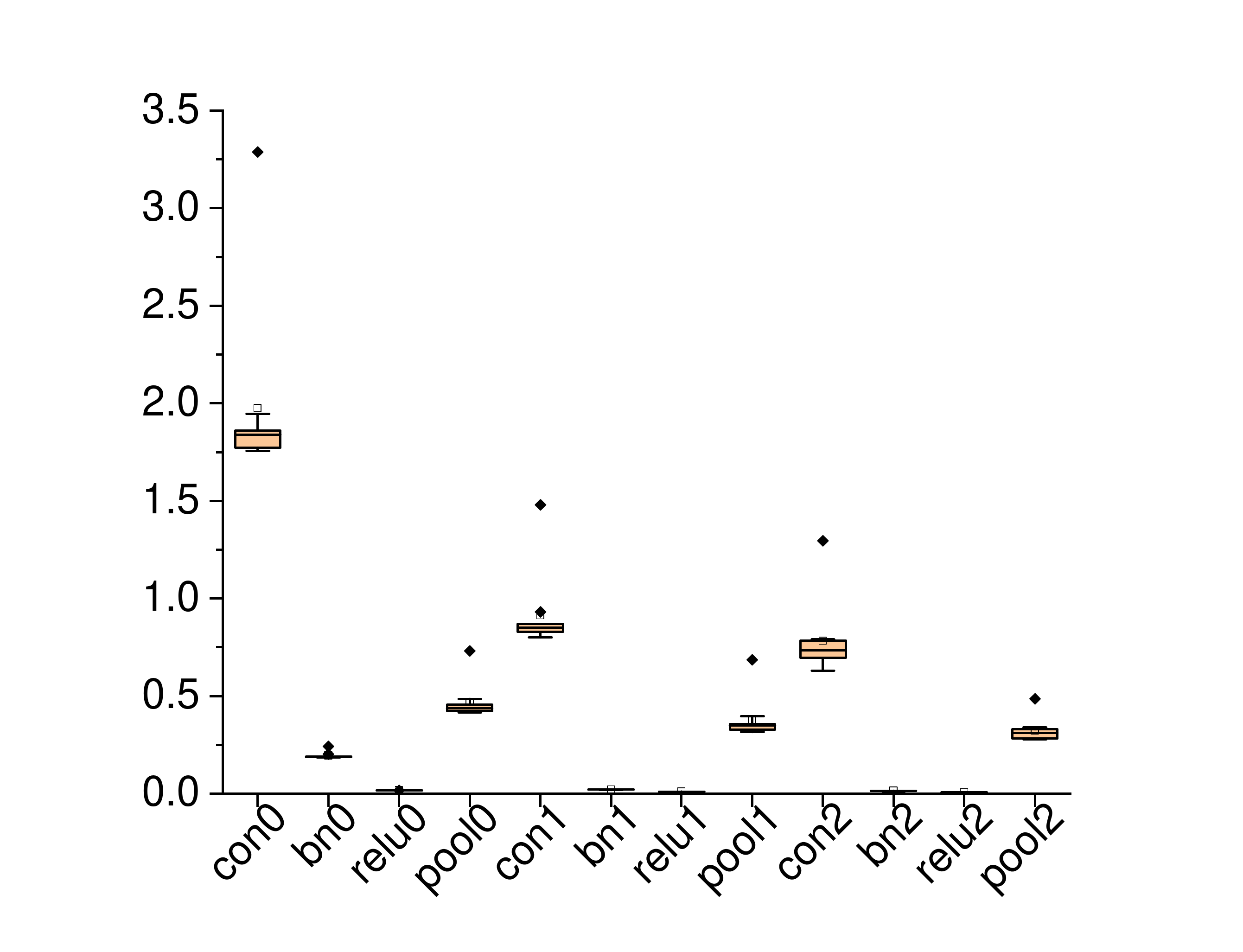} 
           \label{exelatencyofum}
}
\subfloat[Far fault latency (ms).]{
            \includegraphics[width=4.2cm, height = 3.2cm]{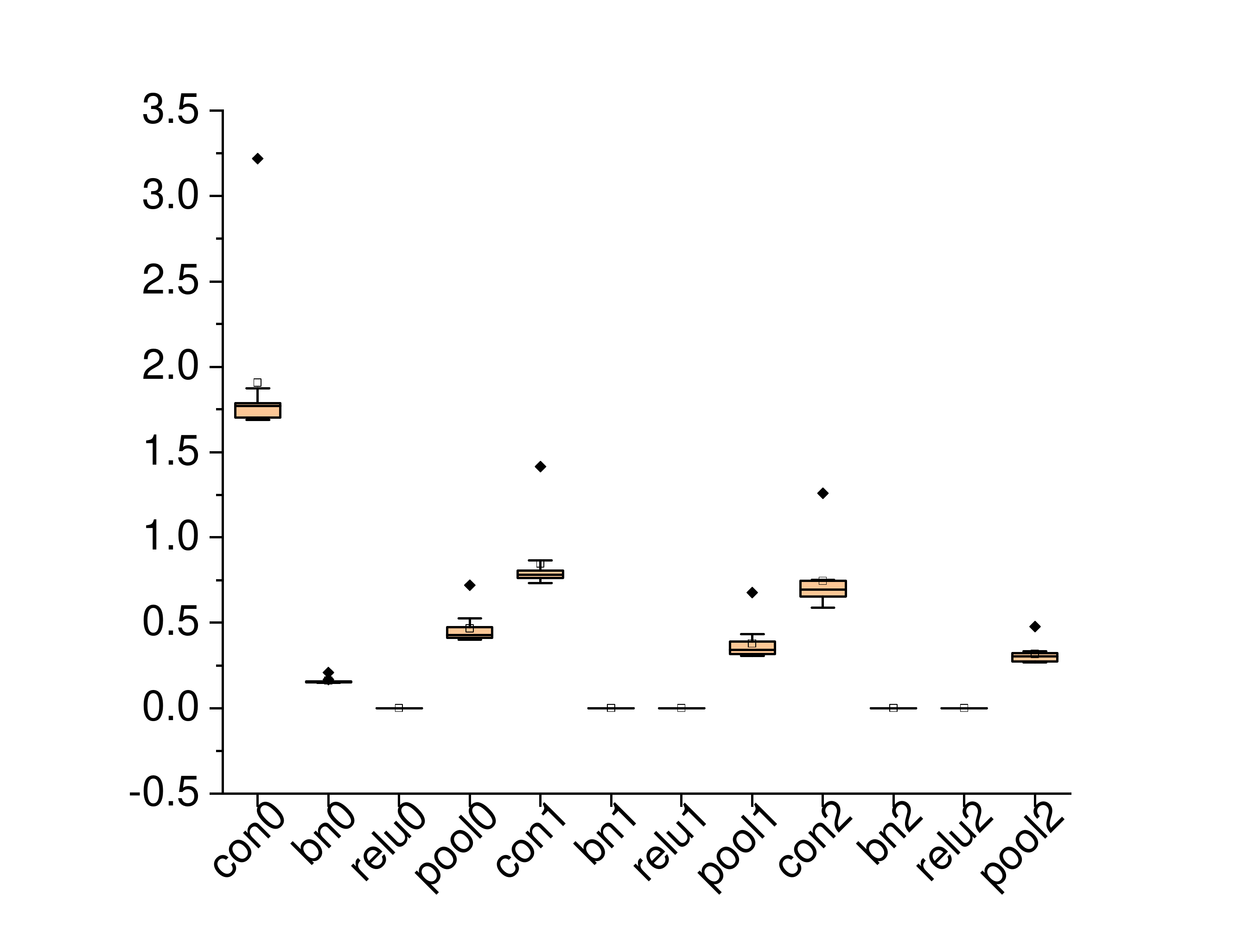}
            \label{PFLatency}
}
\subfloat[Migration latency (ms).]{
            \includegraphics[width=4.2cm, height = 3.2cm]{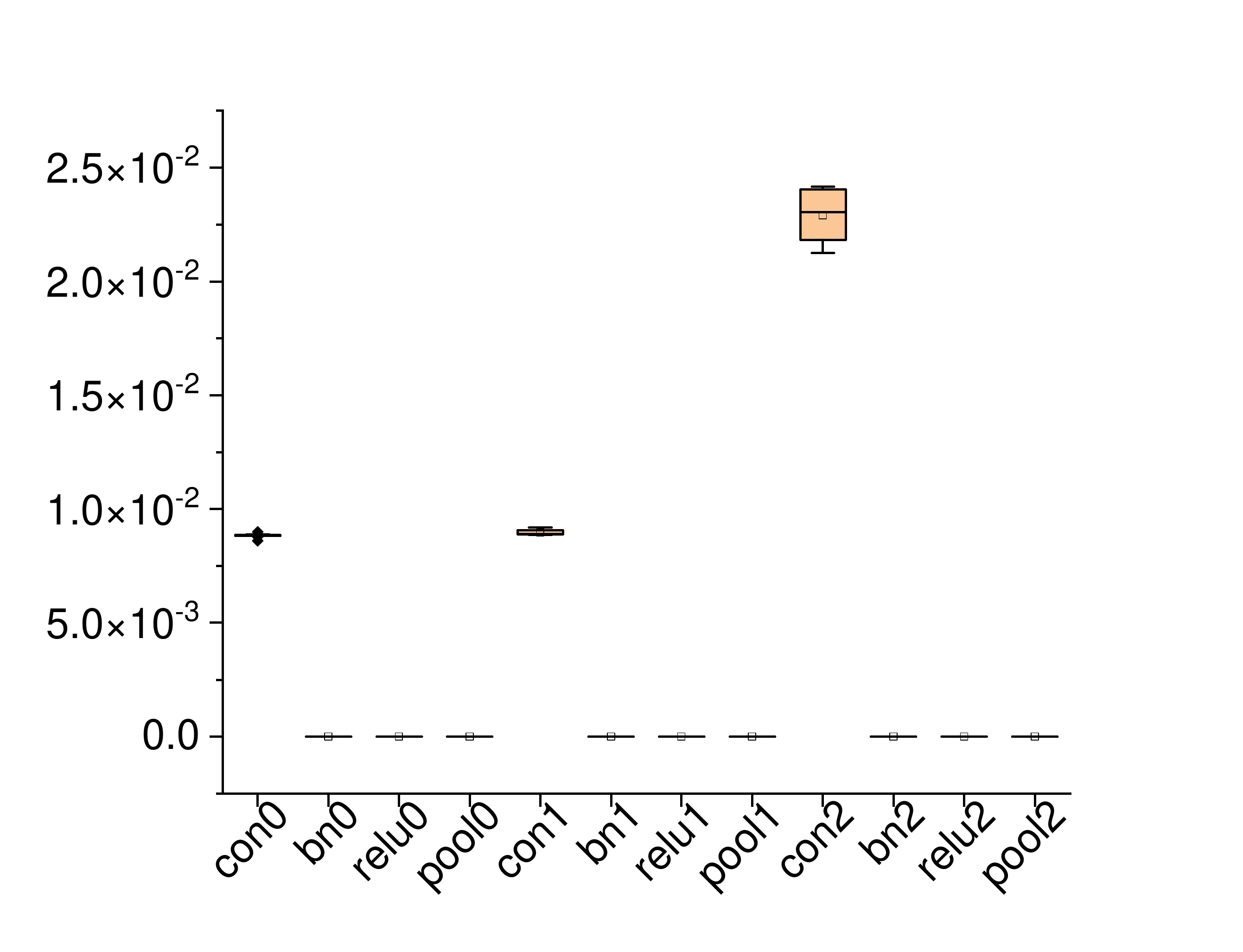} 
            \label{h2d latency}
}
\subfloat[Migration size (KB).]{
            \includegraphics[width=4.2cm, height = 3.2cm]{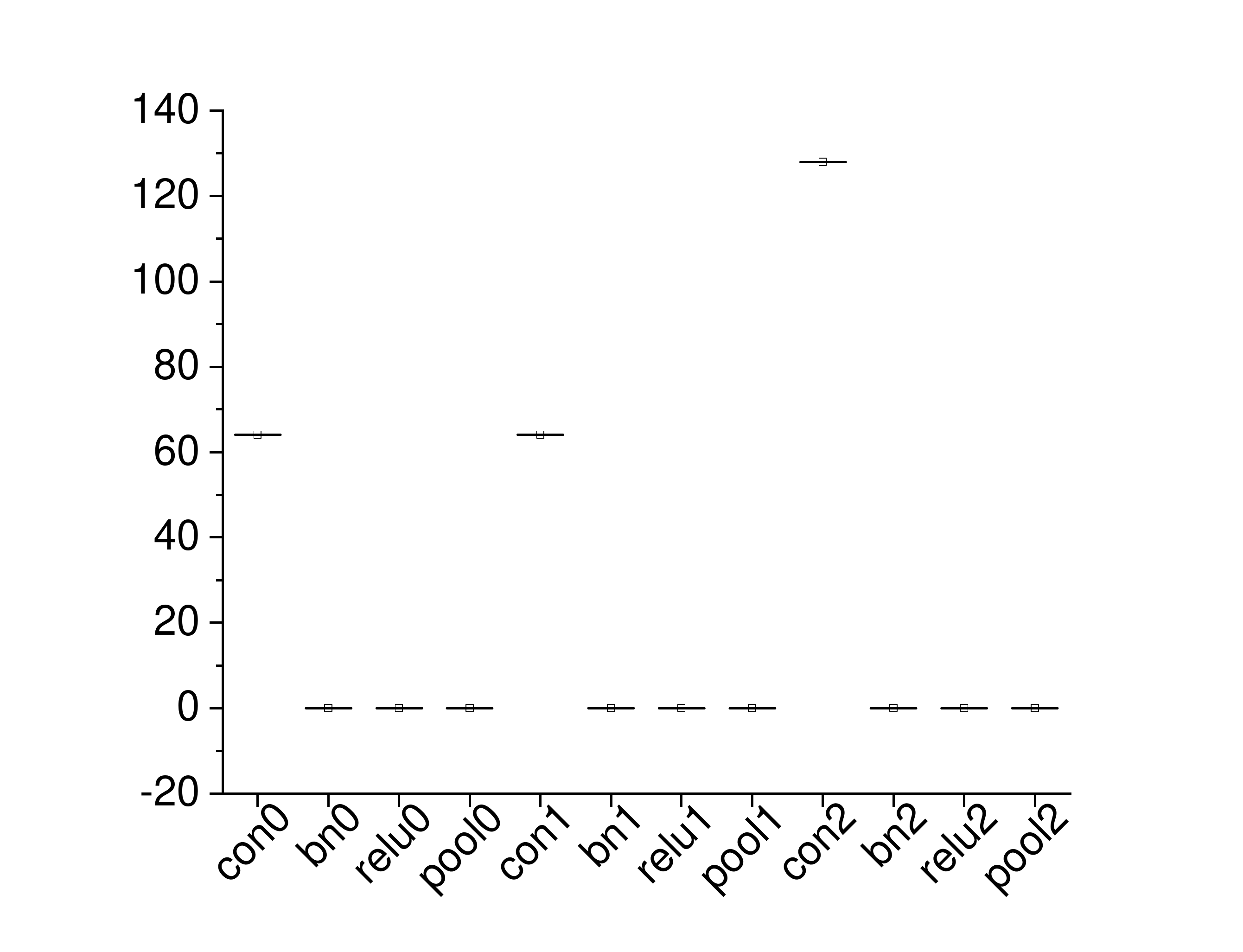} 
            \label{h2dsize}
}
\caption{Comprehensive characterization of distributions of different Arch-hints in UM system.} 
\label{comcharater}
\vspace{-4mm}
\end{figure*}

As discussed in Sec. \ref{archhints}, the ever-explored Arch-hints for GPU platform target copy-then-execute (CoE) system. Since the memory management and data movement in UM system differ from that in CoE, we explore how this difference impacts the Arch-hints patterns and effectiveness in model extraction. 

\noindent\textbf{What Should be Effective Arch-hints:} 
Extraction attack essentially explores the relation between the observed Arch-hints and the internal architectures of the victim model, and typically utilizes a training-based approach to learn the exhibited patterns and leaked information from the Arch-hints to predict the architecture, 
as shown in Table \ref{tb1:AvaiArchhints}. 
For example, \cite{naghibijouybari2018rendered} utilizes the Arch-hints of memory write transactions and unified cache throughput as inputs to train classification models (e.g., KNN) to predict the victim model neuron number. \cite{wei2020leaky} utilizes the Arch-hints, including the number of DRAM read/write requests and the number of texture cache requests, as inputs to train the LSTM model to predict different DNN layer types.

Since the Arch-hints work as the input feature vector of the adversary learning model, the distribution property of Arch-hints across different layers significantly impacts the model extraction performance (e.g., extraction accuracy, Sec. \ref{evalacc}).
It is expected that the Arch-hints distribution across different layers during model execution exhibits clear and accurate patterns. Unfortunately, some Arch-hints distribution can become blurred and inaccurate in UM system. 

\noindent\textbf{Existing Arch-hints May be Ineffective in UM:} 
In fact, we observe that \textit{the CoE platform-targeted Arch-hints can get blurred in UM system during model execution}, which potentially undermines the extraction attack. Typical cases are Arch-hints based on kernel activity or memory traffic.

For example, such a common Arch-hint as kernel latency is closely associted with kernel size. It only consists of kernel execution latency in CoE platform. Since the data has been copied into GPU memory, the kernel execution can access the data in local memory and the execution latency is stable. During model execution, each layer shows stable latency. Due to the different size, different layers show different but stable latencies. The Arch-hint shows clear and accurate distribution across different layers.
In comparison, the kernel latency includes far fault latency and migration latency besides the execution latency in UM \cite{umbeginners}, and the execution latency can overlap with the other two. Thus, each layer shows in-stable and variant latency during model execution, and the Arch-hint shows blurred and inaccurate distribution across different layers. Consequently, the Arch-hint can become ineffective for distinguishing different layers. 

Regarding the memory bus traffic, in CoE platform, as the data has been in GPU memory, the memory read transaction can reveal the input size of a kernel \cite{hu2020deepsniffer}. Thus, the Arch-hint of memory transaction is accurate. However, in UM, besides reading data from DRAM, a kernel can migrate large amount of data from CPU memory on demand, which will be directly loaded into GPU cache. Thus, the Arch-hint of memory transaction is inaccurate to reveal the kernel size. 

Here, we utilize the Darknet reference model as an example to illustrate our observation. We choose Darknet framework in this paper for three reasons: 1) it is open source; 2) it is written in C and CUDA, and well supported with the CUDA UM model APIs (e.g., $cudaMallocManaged()$, $cudaFree()$); 3) it provides a variety of standard pre-trained models for objective classification and detection applications. We execute the model and collect the Arch-hints on GPU platforms with UM system. Besides the commonly-used Arch-hints, we explore three candidate Arch-hints specified in UM model: \textit{far fault latency, data migration latency, and data migration size}, as shown in Table \ref{collectarchhints}.
For each Arch-hint, we execute the model 10 times and collect 10 samples, and then draw the box-plot, as shown in Fig.\ref{comcharater}. 

\noindent\textbf{Quantifying the Effectiveness of Arch-hints:} 
To evaluate how the Arch-hints effectiveness are undermined and the extraction attack performance is impacted, we propose metrics to quantify the effectiveness of each Arch-hint. 
Note that all Arch-hints can be used in extraction attack theoretically, however, an effective Arch-hint can faithfully mirror the patterns of the model execution (e.g., layer features). If the arch-hint is ineffective, it will be more difficult for the adversary to extract the victim model, for example, the adversary has to pay more observations to obtain accurate results.

We mainly evaluate an Arch-hint's effectiveness in terms of its \ul{distribution across the NN layers}. The distribution can be measured using \textit{coefficient of variation (CoV)} from two aspects: 1) $distinguishability$, 2) $consistency$.  
$CoV$ is a statistical measure of the dispersion of a series of data independent of the measurement unit used for the data. As different Arch-hints have different measurement units, $CoV$ is useful for comparing the different Arch-hints distributions. 
$CoV$ is calculated as the ratio of the standard deviation ($\sigma$) to the mean ($\mu$), as shown in Equation \ref{cov}. 
Fig.\ref{comcharater} shows the box-plot of each Arch-hint, where the x-axis indicates the layers in Reference model and the y-axis indicates the 10-samples distribution of the Arch-hint on each layer. We detail below how the $CoV$ is used to measure the $distinguishability$ and $consistency$ of each Arch-hint. Note that we only show early layers of the model to save space, however, the calculation is applied to all layers. 

\begin{equation}
\footnotesize
CoV = \frac{\text{$\sigma$}}{\text{$\mu$}}
\label{cov}
\end{equation}

\textit{a) Distinguishability (dis)} indicates variability of the Arch-hint value among different layers during model execution.
As different layers of a model (e.g., Conv, BN, Pool, etc.) have different computation complexity and dimension size, it is supposed that one Arch-hint behave differently on different layers. 
$distinguishability$ is defined as the variability of the Arch-hint among all layers of a model and is calculated as the $CoV$ of the Arch-hint values of these layers, as shown in Equation \ref{dis}, where the n is total layer number. 
Intuitively, the larger the $CoV_{dis}$, the better the $distinguishability$ is (i.e., the $distinguishability$ positively correlates to $CoV_{dis}$). Then, the larger the Arch-hints difference on different layers is, the easier the adversary can distinguish different layers and explore the model internal architecture, and thus, the more effective the Arch-hint is. 

\begin{equation}
\footnotesize
dis = variability_{Arch-hint}(layer_{1}, layer_{2}, ..., layer_{n})
\label{dis}
\end{equation}


\textit{b) Consistency (con)} indicates the variability of each-layer Arch-hint values among the multiple samples/executions. 
It is expected that the Arch-hint shows consistent behaviors among the multiple samples (i.e., low variability) to provide accurate information about the model architecture, otherwise, the Arch-hint value contain great noises, increasing the difficulty for adversary to accurately extract the model architecture.
As Equation \ref{con} shows, the $consistency$ is calculated as the $CoV$ of each-layer Arch-hint values of the multiple samples, where i $\le$ n. 
Accordingly, the lower the $CoV_{con}$ is, the larger the Arch-hint $consistency$ is (i.e., the $consistency$ negatively correlates to $CoV_{con}$). That is, the more accurate and less-noisy information about the model architecture that Arch-hint can provide, the more effective the Arch-hint is.

\begin{equation}
\footnotesize
con = variability^{layer_i}_{Arch-hint}(sample_{1}, sample_{2}, ..., sample_{m})
\label{con}
\end{equation}


\noindent\textbf{Integratively Evaluate the Effectiveness of Arch-hints:} 
We analyze above that an Arch-hint distribution property among different layers during model execution matters to the Arch-hint effectiveness and the Arch-hint distribution can be measured from both $distinsuishability$ and $consistency$.  
Here, we integrate the $distinsuishability$ and $consistency$ of each Arch-hint, and define the \ul{\textit{Arch-hints Effectiveness Score (ArchES)}} to evaluate the overall effectiveness of an Arch-hint.

The $ArchES$ is defined as \textit{the ratio of distinsuishability to consistency}, that is, $CoV_{dis}$/$CoV_{con}$ (Sec. \ref{effofarchs}). 
On one hand, the $CoV_{dis}$ is expected to be large such that the Arch-hint behaves significantly differently on different layers, providing recognizable information on the model architecture.  
On the other hand, the $CoV_{con}$ is expected to be low such that the Arch-hint behaves consistently among multiple model executions, providing accurate and noise-less information of the model architecture.
The higher $ArchES$ is, the more effective the Arch-hint is. 
By utilizing the $ArchES$, We identify that UM system exhibits several unique and effective Arch-hints, which provides a new attack surface for adversary to extract DNN model (Sec. \ref{newatk}). 

In fact, when we discuss the effectiveness of Arch-hints, we essentially explore whether the data movement during model execution can exhibit distinguishable and accurate patterns, which can be regarded as the leak of the model architecture information, for being learned by adversary in a given system. In conventional Copy-then-Execute (CoE) system, such commonly-used Arch-hints as memory bus traffic, kernel latency can represent the input/output data size and computation complexity of a layer accurately and distinguish different layers. However, they become blurred in UM system. Instead, some new Arch-hints can reveal the data movement pattern clearly and accurately during model execution.

\vspace{-3pt}
\section{Extracting Models with Arch-hints in UM} \label{newatk}
\vspace{-3pt}
In this section, we show how the identified Arch-hints based on page fault handling and on-demand data migration in UM system exhibit patterned information, revealing layer features and DNN characteristics. We then leverage these Arch-hints to launch an extraction attack, termed as \textbf{UMProbe}. To the best of our knowledge, this is the first model extraction attack targets unified memory system. 
\vspace{-3pt}
\subsection{Threat Model and UM Arch-hints} \label{UniqueArchhints}
\vspace{-3pt}

The threat model focuses on edge security where the adversary is able to physically access the victim platform. Also, with GPU multi-instances technology \cite{MIGPU} and GPU's support for the concurrency of multi-tenant inference applications in edge scenarios \cite{liang2020ai}, the adversary can share the physical GPU platform with the victim and co-locate its application with the victim model in the GPU.  

First, the adversary can utilize the PCIe bus snooping method to obtain the GPU kernel and data migration activities, which has been proven with an accuracy of $\sim$98\% in practice\cite{zhu2020hermes}.  
Specifically, the GPU activities are initialized and terminated from the host commands, which are transferred through the PCIe connection between GPU and host. Accordingly, the far-fault handling requests and on-demand page migration both cause PCIe traffic in UM system, as shown in Fig.\ref{umprin}. 
By capturing these critical traffic and events related to far-fault requests and data migration \cite{PCIAnalyzer}, the adversary can obtain the Arch-hints of \textit{Page Fault Latency (PFLat), Page Migration Latency (MigLat) and Migration Size (MigSize)}
of each kernel and layer. We consider these Arch-hints to be \textbf{Primary Arch-hints ($PriArchs$)} in UM system.

We show that the primary Arch-hints can show strong patterns and are effective in leaking model information for adversary and that UMProbe is able to extract the victim DNN architecture accurately by merely leveraging these $PriArchs$. 

Additionally, since the adversary and victim share the same GPU platform (e.g., the hardware cache/memory, the open deep learning library (e.g., Darknet)), as demonstrated in \cite{naghibijouybari2017constructing, naghibijouybari2018rendered}, the adversary can co-locate its spy application with the victim model to obtain the victim cache properties (e.g., L2 transactions) by leveraging cache-side channels, which can achieve $\sim$ 90\% accuracy. 
UMProbe can collect the \textit{common Arch-hints ($ComArchs$), such as L2 read and write transactions} to further enhance its extraction performance. 




\begin{figure}[t]
\centering
            \includegraphics[width=\linewidth]{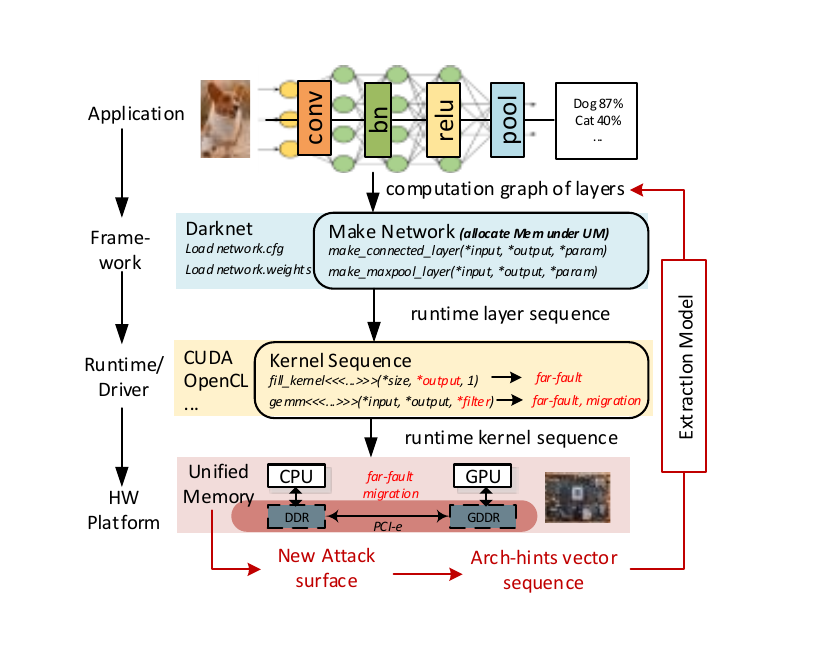}
            \caption{Overview of UMProbe.}
            \vspace{-2mm}
            \label{dataflowofumprobe}
\end{figure}

\noindent\textbf{UMProbe Overview:} After obtaining the Arch-hints, we overview how $PriArchs$ exhibit patterned information to reveal different layer features and manifest model architecture. 

As show in Fig.\ref{dataflowofumprobe}, a DNN application issues its computation graph to the Darknet framework, and Darknet forms the graph as the connected layer sequence of the DNN architecture. Then, Darknet transforms the layer sequence into the GPU commands of runtime kernel execution sequence corresponding to the layer sequence. Finally, the kernel sequence is executed in GPU platforms, which exhibits the Arch-hints of being learned by the adversary.

Regarding Darknet framework, it mainly loads the network configuration file and parameters file (i.e., weights) after receiving the DNN execution request. Essentially, Darknet constitutes the network architecture and allocates memory space for each layer (i.e., the IFM, OFM, Filter) in UM system by calling the API $cudaMallocManaged()$. 
In UM system, it is lazy allocation, indicating that the physical page of the data populates in the host memory and the virtual page is invalid in the GPU side (i.e., the virtual-to-physical mapping does not exist or page valid flag is not set) after allocation completes. Thus, when the GPU SMs execute the layer and kennel sequentially, the SMs will encounter far-page fault exception if the SMs access the data region for the first time and may cause on-demand page migration. 

As different layer types utilize different GPU kernels, the $PriArchs$ can exhibit patterned information in leaking the different kernels characteristics and layers features.  
Considering the kernel sequences of different layers have a static execution order related to the original computational graph of a DNN model \cite{hu2020deepsniffer}, the kernels characteristics and layers features revealed by $PriArchs$ can be learned by adversary to predict the model layer sequence accurately.

  
\vspace{-3pt}
\subsection{New Attack Surface in UM}\label{atksurface}
\vspace{-3pt}
Essentially, UMProbe utilizes a learning-based model to explore the relationship between the extracted Arch-hints and victim model's internal architectures, the input Arch-hints containing the victim architecture information can reveal the victim layer features. In this section, we show how the primary Arch-hints under UM represent the different kernels' characteristics and reveal different layer features, which thus exposes a new attack surface in unified memory system for the adversary to infer victim DNN architecture.


\subsubsection{Primary Arch-hints Reveal Layer Features}\label{unireveal}
\noindent\textbf{Primary Arch-hints Vary with Layer OFM/Filter Characteristics:}
As a DNN layer can be specified by its feature map (i.e., IFM, OFM) and its parameters (e.g., the Filter of a Conv layer), we observe that the primary Arch-hints of \textit{PFLat, MigLat and MigSize} are closely associated with the feature map and parameter characteristics of a runtime layer. 
Table \ref{tb2:kerlayfeature} shows the most-commonly-used layer types in a DNN model. By analyzing Darknet code, we identify the associated kernels of each layer. We will analyze how these kernels behave during DNN execution and how the Arch-hints can eventually reveal the kernel characteristics and layer features.

\begin{table}[t]
\centering
\scriptsize
\caption{Associated kernels and Arch-hints of the typical layers in Darknet.}
\vspace{-6pt}
\begin{tabular}{l|l|c|c|c}
\hline
\tabincell{l}{Layer} &\tabincell{l}{Kernels} &\tabincell{l}{PFLat} &\tabincell{l}{MigLat} &\tabincell{l}{MigSize}\\\hline
\tabincell{l}{Conv} &\tabincell{l}{fill\_kernel, im2col\_kernel, \\gemmSN\_kernel\_nn, sgemm\_xx\_nn}&\tabincell{c}{\checkmark} &\tabincell{c}{\checkmark} &\tabincell{c}{\checkmark}\\\hline
\tabincell{l}{FC} &\tabincell{l}{fill\_kernel, gemmSN\_kernel\_tn, \\sgemm\_xx\_tn, axpy\_kernel,} &\tabincell{c}{\checkmark}&\tabincell{c}{\checkmark} &\tabincell{c}{\checkmark}\\\hline
\tabincell{l}{BN} &\tabincell{l}{normalize\_kernel, scale\_bias\_kernel,\\ add\_bias\_kernel} &\tabincell{l}{}&\tabincell{l}{}&\tabincell{l}{}\\\hline
\tabincell{l}{ACT} &\tabincell{l}{activate\_kernel (ReLu)}&\tabincell{l}{} &\tabincell{l}{} &\tabincell{l}{}\\\hline
\tabincell{l}{Pool} &\tabincell{l}{forward\_maxpool\_kernel, \\ forward\_avgpool\_kernel} &\tabincell{c}{\checkmark} &\tabincell{l}{} &\tabincell{l}{}\\\hline
\tabincell{l}{Shortcut} &\tabincell{c}{copy\_kernel, shortcut\_kernel}
&\tabincell{l}{\checkmark} &\tabincell{l}{} &\tabincell{l}{}\\\hline
\end{tabular}
\vspace{-6mm}
\label{tb2:kerlayfeature}
\end{table}

\textit{a) Conv and FC layers.} 
The execution of a Conv layer involves multiple kernels, such as $fill\_kernel$, $gemm\_kernel$. Here, the kernel $fill\_kernel$ works to initialize the OFM region of the layer (i.e., filling value $1$ in the region), which is allocated in GPU memory before the convolution operation (i.e., the kernel $gemm\_kernel$) begins. To initialize the region, the GPU SMs access it for the first time after the memory being allocated, causing far-page fault handling and PFLat. 

After the OFM region is initialized, the kernels $im2col\_kernel$ and $gemm\_kernel$ are launched to execute convolution operation. During $gemm\_kernel$ execution, the GPU SMs has to access the Filter region (i.e., storing the weights) for the first time, causing far-page fault handling and PFLat as well. Moreover, since the physical pages of Filter data populate in the remote system memory at this moment, data migration is required after the far-page fault is processed, which results in MigLat and MigSize. 


For the FC layer, it almost follows the same pattern of Conv layer as they are both linear operation layers in a model. Specifically, FC layer starts with the kernel $fill\_kernel$ that initializes the OFM region and can cause far-page fault, and then executes the computation kernel $gemmSN\_kernel\_tn$ that accesses the Filter region and causes both far-page fault and data migration, and ends with the kernel $axpy\_kernel$ that again accesses OFM region and does not cause far-page fault or migration.  

Besides, although the layer implementations may vary in runtime, like a Conv layer can be implemented with a $gemmSN\_kernel\_nn$ or a $sgemm\_xx\_nn$ kernel (xx indicates different dimension, such as $32\times32$, $64\times32$), this makes no difference to our analysis above.
These kernels always access the OFM and Filter region and result in both far-page fault handling and data migration.


\textit{b) BN and ACT layers.}
A Conv layer is typically followed by a BN layer and then a ACT layer. As Table \ref{tb2:kerlayfeature} shows, BN layer consists of three kernels, including $normalize\_kernel$, $scale\_bias\_kernel$, and $add\_bias\_kernel$, and ACT layer consists of $activate\_kernel$ (e.g., the most-commonly-used ReLU). 
When SMs execute a BN layer, the layer takes the OFM of the previous layer as its own IFM, indicating the OFM region has been accessed before and the data pages have populated in the local GPU memory. Thus, the SMs accessing the region does not cause far-page fault or data migration. Analogously, the ACT layer execution causes no far-page fault or migration. 

Besides, some earlier DNN models, such as Alexnet, do not include BN layer, the ACT layer directly takes the previous Conv/FC layer OFM as its IFM. Also, modern models include BN and ACT layers in Conv layer, known as Conv-BN-ReLu block \cite{ioffe2015batch}. However, our analysis above is also applicable to these different models variants.
Namely, both BN and ACT layers do not cause far-page fault or data migration.



   

\textit{c) Pooling and Shortcut layers.}
Pooling layer mainly involves the kernel $maxpool\_kernel$ or $avgpool\_kernel$, which outputs the down-sampling result of the previous layer. When SMs execute the kernel, the SMs access the OFM region of the Pooling layer for the first time, causing far-page fault handling and PFLat. During the down-sampling operation, the SMs does not need other parameters \cite{cnnparas} and does not cause data migrate. Thus, the Pooling layer is featured with far-page fault handling but no data migration. 

Modern DNN models can be configured with more complex non-sequential architecture, such as the popular ResNet using shortcut connection. During runtime execution, the shortcut and the main branch are actually executed in sequence in GPU platforms \cite{hu2020deepsniffer}. 

In Darknet, the shortcut layer is composed of three kernels, including $copy\_kernel$, $shortcut\_kernel$, $activate\_kernel$. The kernel $copy\_kernel$ is first executed to copy the identity of the IFM of the divergence point to the OFM region of shortcut layer, then the kernel $shortcut\_kernel$ is executed to perform addition operation in OFM region again, and finally the kernel $activate\_kernel$ is executed. As Pooling layer does, the shortcut layer accesses its OFM region for the first time during kernel $copy\_kernel$ execution and causes far-fault handling, however, it does not require additional parameters in layer execution, thus causing no data migration. 

\begin{figure}[t]
\subfloat
{
\centering
           \includegraphics[width=3.6cm, height = 2.5cm]{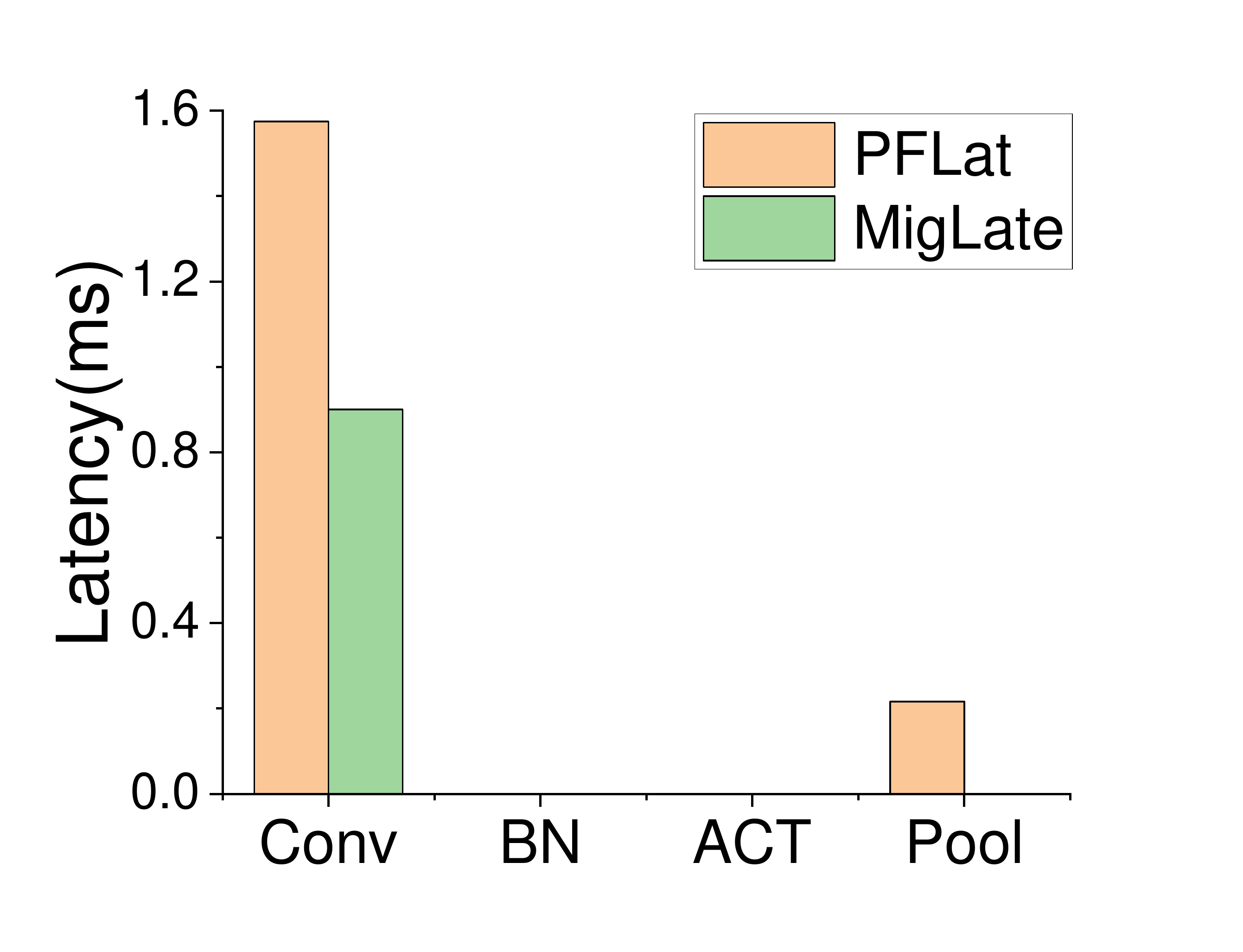}
            \label{reflay}
}
\subfloat
{
            \includegraphics[width=4.3cm, height = 2.5cm]{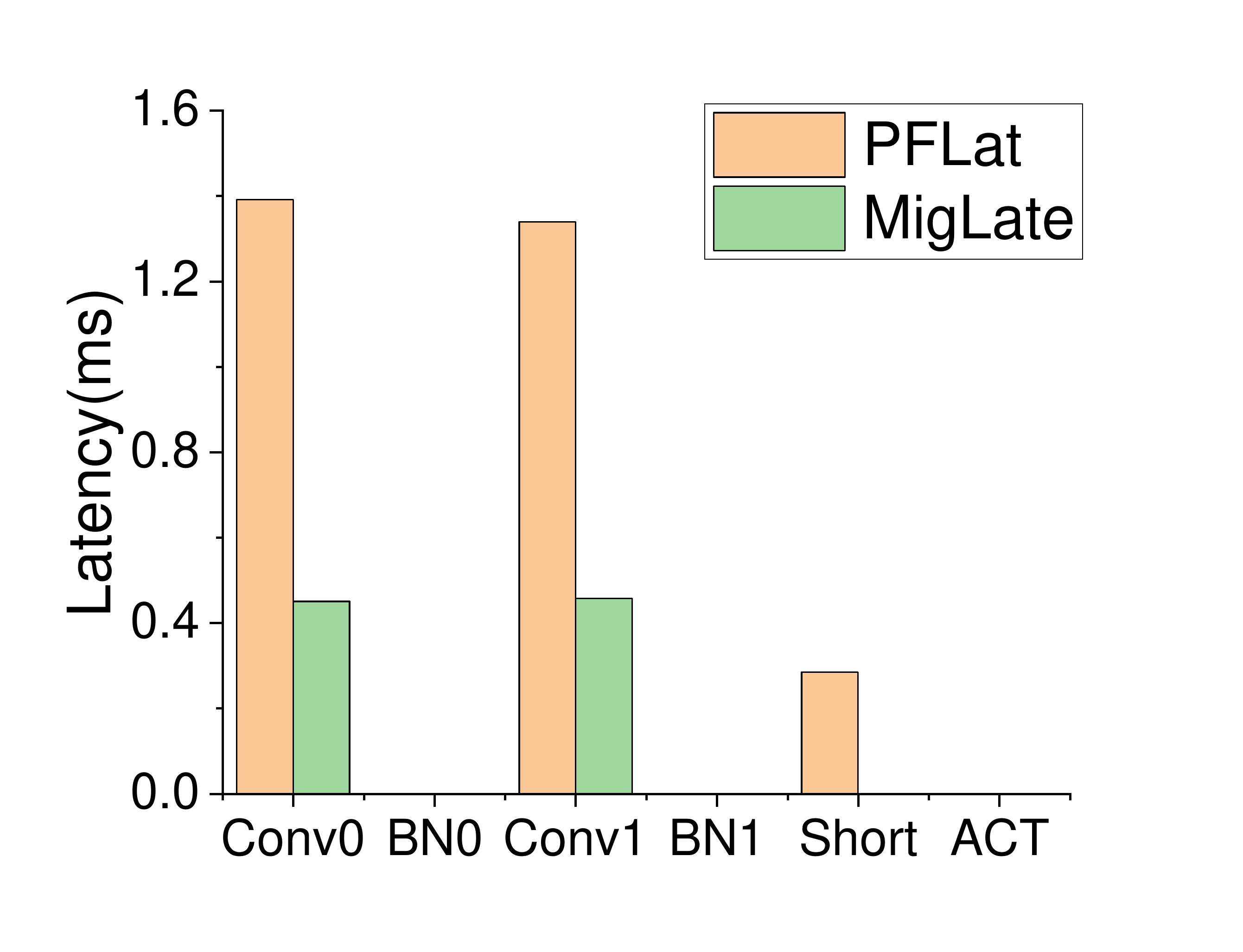}
            \label{reslay}
}
\vspace{-2mm}
\caption{Layer feature from one block of a) Reference, b) ResNet18. For Conv layer, the migration causes MigSize 4608B for Reference and 2304B, 2304B for ResNet18, respectively.}
\label{layerfeaturedata}
\vspace{-4mm}
\end{figure}

\textbf{Observation-1:} During DNN model execution, different types of layers perform differently and cause different patterns on page fault handling and on-demand data migration, as shown in Table \ref{tb2:kerlayfeature}. 
The Conv/FC layer is featured with both \textit{PFLat, MigLat and MigSize}, while the layers of BN and ACT do not cause far-page fault or data migration. Both the Pooling and Shortcut layers only cause far-page fault and \textit{PFLat}. 
Fig.\ref{layerfeaturedata} shows examples of one Reference block and one ResNet18 residual block, we observe their primary Arch-hints behave as we discussed above.

\noindent\textbf{Primary Arch-hints Reveal Filter Size:}
As the Conv/FC layer is featured with the Arch-hints of far-page fault and data migration, and the data migration is mainly caused by the Filter data of the layer, we explore how the primary Arch-hints of \textit{PFLat, MigLat, MigSize} can reveal the Filter Size characteristic.  
As Conv layer is the dominant layer in DNN architecture, we characterize all the Conv layers of Reference model to show the analysis.

\begin{figure}[t]
\centering
            \includegraphics[width=8.3cm, height = 3.2cm]{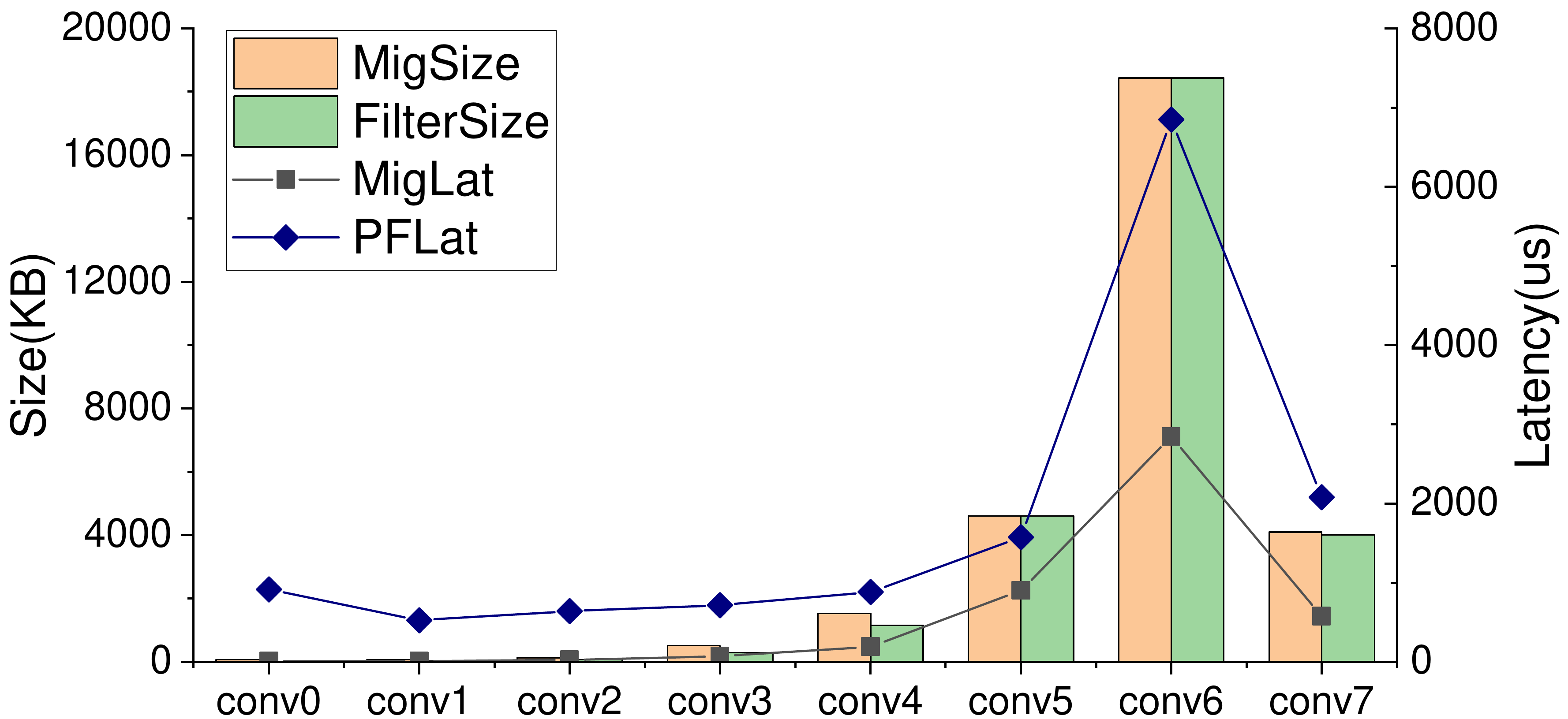}
            \vspace{-2mm}
            \caption{Arch-hints reveal Filter size in Reference model.}
            \label{paraamountforarchs}
            \vspace{-5mm}
\end{figure}

The Filter size of a Conv layer can be calculated as $Channel_{IFM}$ $\times$ $Width_{Filter}$ $\times$ $Height_{Filter}$ $\times$ $Channel_{OFM}$ $\times$ 4 bytes (i.e., each data is a $float$ type in memory).
As shown in Fig.\ref{paraamountforarchs}, the x-axis indicates the different Conv layers with the network going deeper, and the y-axis on left-hand side indicates the migration and Filter data size, and the y-axis on the right-hand side indicates page fault and migration latency.  
We observe that the MigSize is almost equal to the Filter Size, indicating that the migration is mainly caused by the Filter data. Also, with network going deeper, the Filter Size increases, and MigSize increases accordingly. 

Meanwhile, the MigLat and PFLat increases as well, following the trend of Filter Size and MigSize. Intuitively, increasing MigSize causes increasing MigLat. Also, with Filter Size increasing, the SMs have to access a larger Filter data region in the memory during layer execution, which can trigger a larger amount of far page fault latency.

\textbf{Observation-2:} 
During Conv/FC layer execution, the data migration mainly result from the Filter data. Thus, the migration data size well reveals the Filter data size, and the far fault latency and migration latency both positively correlates to Filter data size.

\noindent\textbf{Primary Arch-hints Reveal Layer Features:} 
As the primary Arch-hints of PFLat, MigLat, MigSize can reveal OFM size and Filter size, we show how these primary Arch-hints can leak layer features and manifest model architecture during model execution. 

\begin{figure}[t]
            \includegraphics[width=8.3cm, height = 3.5cm]{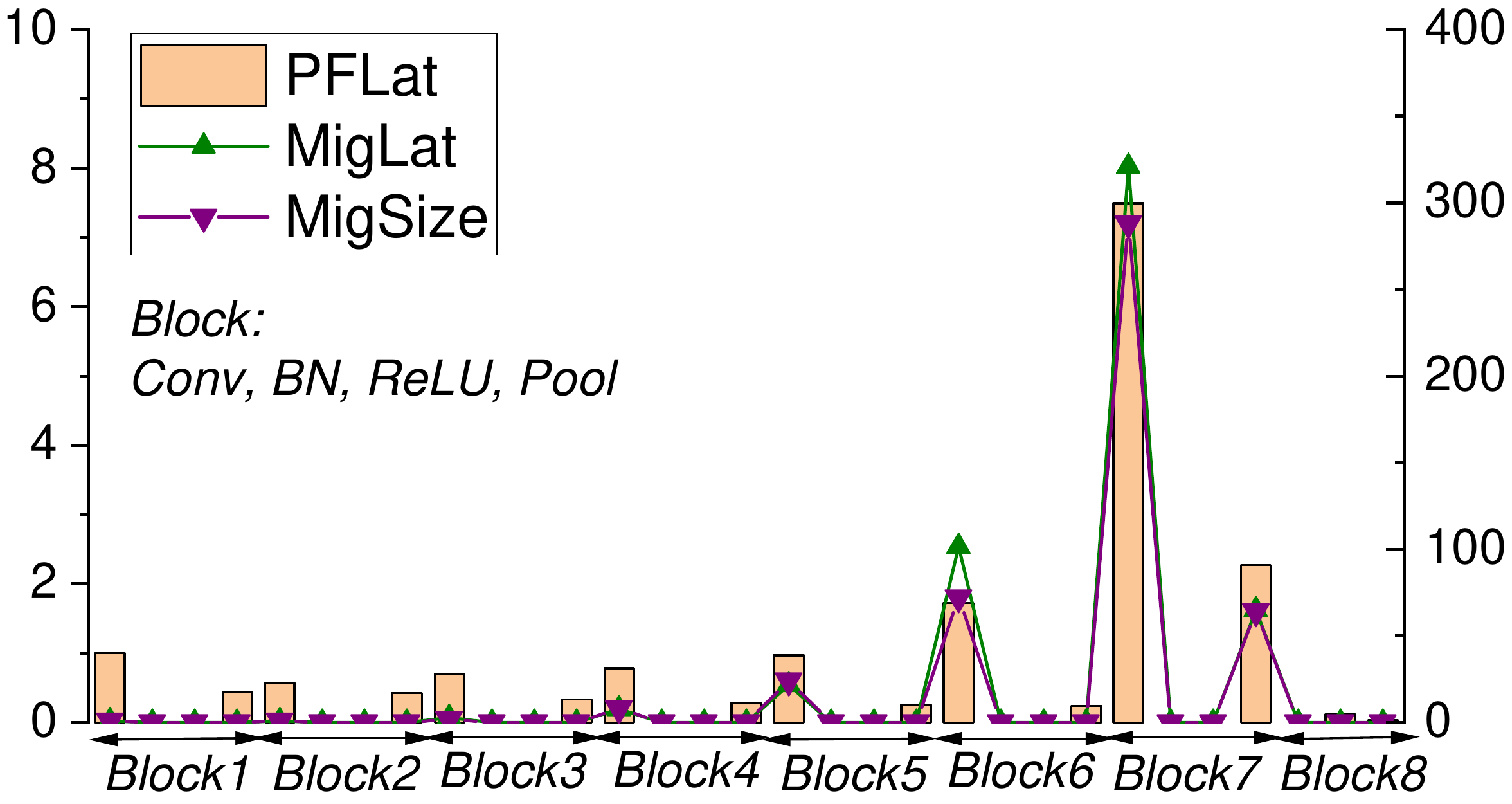}
            \vspace{-2mm}
            \caption{The Arch-hints of different blocks in Reference.}
            \label{difftypesforrefer}
            \vspace{-5mm}
\end{figure}

We characterize the Arch-hints \textit{PFLat, MigLat, MigSize} of each layer (e.g., Conv, Pool, FC) during Reference model execution, as shown in Fig.\ref{difftypesforrefer}. The x-axis in Fig.\ref{difftypesforrefer} indicates different blocks as the network going deeper, and the y-axis on the left hand indicates the latency while the right hand indicates the data size. 

We observe that as the network goes deeper, the scale of \textit{PFLat, MigLat, MigSize} increases significantly, especially for the Conv layer. The different scales can identify the different blocks. 
Second, the \textit{PFLat, MigLat, MigSize} of a Conv layer are usually much larger than other types of layers (e.g., Pool, the last layer) within the same block. This is because the large Filter size and feature map size in a Conv layer cause large amounts of far page fault and data migration. 
Third, the BN layer and ACT layer (i.e., ReLU) exhibit quite similar execution features as they both do not cause far page fault or migration, resulting in difficulties for the adversary to accurately distinguish them. 


\textbf{Observation-3:} 
The primary Arch-hints of \textit{PFLat, MigLat, MigSize} can reveal different types of layers and blocks during model execution by identifying the layer's feature map and Filter characteristics and leak information on the model internal architecture. 
Thus, these Arch-hints exposes a new attack surface in UM system for extraction attack, which has not been explored before.

\noindent\textbf{Common Arch-hints Further Helps:}\label{l2cache}
We analyze above the primary Arch-hints of \textit{PFLat, MigLat, MigSize} can leak the layer information during model execution. However, some adjacent layers, like BN and ACT, do not cause \textit{PFLat, MigLat, MigSize}, and thus exhibit similar execution features, causing difficulties to UMProbe. Although UMProbe can utilize DNN model design philosophy (i.e., the empiric to follow a BN and ACT layer after Conv layer) to infer these layers, we consider UMProbe exploring other Arch-hints in UM system to conquer the difficulties. 

As we analyzed in Sec. \ref{archhintcharacter}, the L2 read/write transaction obtains a high $ArES$ and is considered effective in extraction attack.  
Thus, UMProbe utilizes the common Arch-hints of L2 read/write transaction in the attack besides the primary Arch-hints of \textit{PFLat, MigLat, MigSize}. 
Fig.\ref{l2write} and \ref{l2read} shows the Arch-hint of L2 write/read transaction shows noticeable difference on the BN and ACT layers, indicating UMProbe can utilize the Arch-hints to further improve its extraction accuracy (Sec. \ref{evalacc}).

\subsubsection{Learning-based Extraction Attack} \label{atk}
\vspace{-3pt}
As We learned that the Arch-hints in UM system, especially the primary Arch-hints, can reveal layer features and leak model information during model execution, we will show how UMProbe can extract and identify victim model by learning these Arch-hints. 
Since model architecture, especially model layer sequence, is the most fundamental one among DNN model's properties and can be used to infer other parameters \cite{tramer2016stealing, wang2018stealing, liu2016delving, oh2019towards, hu2020deepsniffer, hu2021systematic, hu2020deepsniffer}, UMProbe is designed to identify the model architecture as the first step to extract the model.


\noindent\textbf{Attack Methodology:} UMProbe adopts the Connectionist Temporal Classification (CTC) \cite{graves2006connectionist} model to predict victim layer sequence including layer number, types and connection, which has been proven effective in \cite{hu2020deepsniffer}. 
CTC is a sequence-to-sequence model, and it can be trained by minimizing the difference between the ground-truth layer sequence $L^{\ast}$ and predicted layer sequence $L$, and outputs a layer sequence which is as close to the ground-truth as possible. 

\begin{figure}[t]
\centering
            \includegraphics[width=8cm, height = 5.5cm]{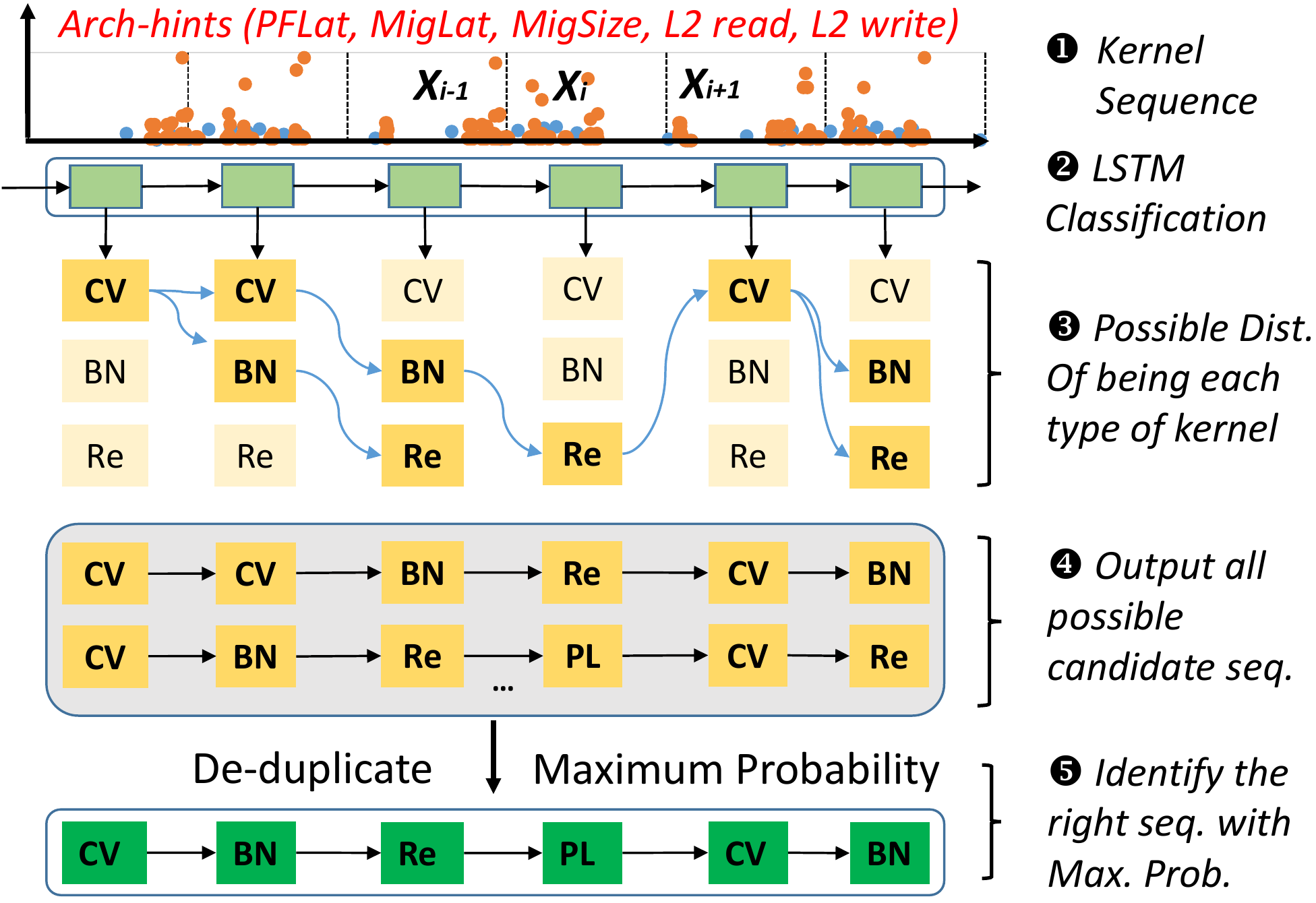}
            \vspace{-3mm}
            \caption{Scheme of layer sequence prediction in UMProbe.}\label{design}
            \vspace{-14pt}
\end{figure}

Fig.\ref{design} shows the specific attack methodology that includes 5 steps. \ding{172} The kernel sequence is composed of multiple kernels featured with their own Arch-hints Vectors $X_{i}$ (i.e., <\textit{PFLat, MigLat, MigSize, L2 read, L2 write}>). 
\ding{173} For the $i_{th}$ kernel, its Arch-hints $X_{i}$ can reveal the characteristics of the kernel. Then, UMProbe conducts the $i_{th}$ kernel classification based on $X_{i}$ by using a LSTM-classification model \cite{hu2020deepsniffer} and \ding{174} will output a probability distribution $K_{i}$ of which type of layer(i.e., Conv, ReLU, BN, Pool, etc.) those kernels belong to.
\ding{175} UMProbe utilizes the CTC model to estimate the conditional probability with the distribution of prior kernels(i.e., $K_{1}$, $K_{2}$, ... $K_{i}$). Then UMProbe outputs all of the kernel sequence candidates here, such as (CV-CV-BN-Re), (CV-BN-Re-PL), etc. 
\ding{176} The CTC decoder eventually recognize the kernel sequence with the largest conditional possibility as the output $L$ by utilizing greedy search and de-duplication techniques \cite{zenkel2017comparison}.
Table \ref{tb2:kerlayfeature} shows each layer is associated with their own specific kernels, and Fig.\ref{dataflowofumprobe} shows a DNN layer sequence that can be mapped to the kernel sequence in runtime, thus, UMProbe can successfully predict the layer sequence by extracting and identifying the kernel sequence.

With layer sequence predicted, we then show how the layer dimension is estimated, though \cite{hu2020deepsniffer} demonstrates that the layer dimension is less important than layer sequence in extraction attack.
Specifically, \cite{hu2020deepsniffer} provides a method utilizing the DRAM read transaction to estimate the input and output size of ReLU layer and other layers.
Similarly, UMProbe utilizes L2 read transaction to estimate the input and output size of different layers by following the same method. 
Regarding GPU memory hierarchy, L2 cache read transaction can provide more accurate information to estimate the input and output size during kernel execution as the L2 cache cannot be bypassed in kernel transaction. Moreover, as we analyzed in Sec. \ref{atksurface} that the MigSize can reveal the Filter size of a layer (i.e., Conv/FC). 
That is, in UM system, the new attack surface, especially the Arch-hint of MigSize, has a advantage in estimating the Filter size of a layer. 
\vspace{-3pt}
\section{Evaluation}\label{evaluation}
\vspace{-3pt}
\subsection{Experimental Setup}
\vspace{-3pt}
\noindent\textbf{Platform:}
All sample collection, model training and validation, and attack evaluation are conducted on NVIDIA Titan RTX GPU platform. The DNN models are implemented in Darknet framework, with CUDA 10.0. 
We use the GPU performance counter \cite{CUPTI} to emulate bus snooping for page fault latency, page migration latency, page migration size and L2 cache read/write transaction information collection.

\noindent\textbf{Benchmarks:}\label{bench}
We use multiple pre-trained DNNs on Darknet framework \cite{darknetclassification}. The benchmark includes Sequential models (Alexnet, VGG-16, Reference, Tiny Darknet, and Extraction \cite{darknetclassification}) and Non-Sequential models (Resnet18, Resnet50, and Resnet101 \cite{he2016deep}). 
It is important to emphasize that, all of the aforementioned models do {\it not} have specific corresponding UM implementations in public domain. We substantially modify the Darknet framework to support its execution in UM system using CUDA APIs $cudaMallocManaged()$ and $cudaFree()$.

\noindent\textbf{Model training and deployment:}
Essentially, UMProbe contains a LSTM+CTC learning model to extract the victim DNN architecture.  
To train UMProbe, we randomly generate enough numbers of DNN models (i.e., random layer number, types, connections and dimensions) with both sequential and non-sequential connections, and utilize them as white-box models. 
We then execute the DNN models and collect the kernel execution samples (i.e., the Arch-hints of DNN kernel sequence) as the input to the model. After model being trained, we test UMProbe on the representative DNN benchmarks. These DNNs work as black-box models to UMProbe, and UMProbe predicts their model architectures by analyzing their Arch-hints exposed.

We collect five types of samples to train/test UMProbe, as shown in Table \ref{archhintsample}, that is, samples using Arch-hints of 1) PFLat, 2) MigSize, 3) PFLat, MigLat, MigSize (PriArchs), 4) L2 read transaction, L2 write transaction (ComArchs), 5) PFLat, MigLat, MigSize, L2 read transaction, L2 write transaction (AllArchs). 
As different Arch-hints represent different DNN model characteristics, we evaluate the UMProbe performance by using different Arch-hints (Sec. \ref{evalacc}).



\begin{table}[t]
\small
\centering \setlength \tabcolsep{1pt}
\caption{Effectiveness Evaluation of Arch-hints in UM.}\label{evaarchs}
\vspace{0pt}
\begin{tabular}{l|ccc}
\toprule
Arch-hints  & Distinguishability/$CoV_{dis}$ & Consistency/$CoV_{con}$ & ArES \\\bottomrule
\tabincell{l}{L2 write trans.}  & \tabincell{l}{1.55}  & \tabincell{l}{0.0017}  & \tabincell{l}{873.41}\\ \hline
\tabincell{l}{DRAM write trans.} & \tabincell{l}{1.56} & \tabincell{l}{0.22} &\tabincell{l}{6.84} \\\bottomrule
\tabincell{l}{L2 read trans.}   &\tabincell{l}{1.71} &\tabincell{l}{0.46} & \tabincell{l}{3.72}\\\bottomrule
\tabincell{l}{DRAM read trans.}   &\tabincell{l}{1.34} &\tabincell{l}{0.51} & \tabincell{l}{2.62}\\\bottomrule
\tabincell{l}{Kernel latency}   &\tabincell{l}{1.81} &\tabincell{l}{0.11}& \tabincell{l}{16.59}\\\bottomrule
\tabincell{l}{Far fault latency}   &\tabincell{l}{2.24} &\tabincell{l}{0.095}& \tabincell{l}{23.38}\\\bottomrule
\tabincell{l}{Migration latency}   &\tabincell{l}{3.58} &\tabincell{l}{0.012} & \tabincell{l}{293.84}\\\bottomrule
\tabincell{l}{Migration size}   &\tabincell{l}{3.55} &\tabincell{l}{0.0081} & \tabincell{l}{437.14}\\\bottomrule
\end{tabular}
\vspace{-5mm}
\label{AvES}
\end{table}

\subsection{Effectiveness of Different Arch-hints}\label{effofarchs}
\noindent\textbf{Metric:} As characterized in Sec. \ref{archhintcharacter}, we define \ul{\textit{Arch-hints Effectiveness Score (\textit{ArchES})}} to quantify the effectiveness of each Arch-hint in UM system in terms of {\it distinguishability} ($CoV_{dis}$) and $consistency$ ($CoV_{con}$), as shown in Equation \ref{score}.
Regarding the Arch-hint, the higher the $distinguishability$ is (i.e., higher $CoV_{dis}$) and the stronger the $consistency$ is (i.e., lower $CoV_{con}$), the more effective the Arch-hint is. 

\begin{equation}
\footnotesize
effectiveness\_score = \frac{\text{$CoV_{distinsuishability}$}}{\text{$CoV_{consistency}$}} = \frac{\text{$CoV_{dis}$}}{\text{$CoV_{con}$}}
\label{score}
\end{equation}

\noindent\textbf{Evaluation:} We then calculate the $ArchES$ of each Arch-hint as well as their $dis$ and $con$ factors, as shown in Table \ref{AvES}. 
We observe that the Arch-hints of L2 write trans, migration latency and size gain much higher $ArchES$ than the other Arch-hints due to the high $dis$ and strong $con$. 
When comparing the L2 transaction to the DRAM transaction, we find that their $dis$ is competitive, however, the $con$ of L2 write transaction is significantly lower than that of DRAM write transaction. Because of the limited capacity of L2 cache, the data is eventually written back to the DRAM. Thus, the total amount of L2 write transaction is typically capped by the L2 cache capacity and shows strong consistency. In comparison, the DRAM capacity is much larger, and there is little limit to the DRAM write transaction, thus, DRAM write transaction shows much more inconsistency. Accordingly, the L2 read transaction shows more consistency than DRAM read transaction.  

Then, taking the remaining Arch-hints into consideration, the migration latency and migration size obviously outperform the other two in terms of $ArchES$. The kernel latency shows the lowest $ArchES$ compared to far fault latency due to its low $dis$ (i.e., low $CoV_{con}$).
This is because, in UM system, the kernel latency is composed of execution latency, far fault latency and migration latency, and the execution latency can overlap wit far fault latency, causing overall kernel latency variable \cite{umbeginners}. 
Thus, the kernel latency can get blurred in multiple execution and the Arch-hint of far fault latency can be more effective.

To summarize, we learned that in UM system the Arch-hints of L2 write transaction, L2 read transaction, far fault latency, and migration latency and size show greater effectiveness in terms of $ArchES$ compared to the other Arch-hints. Essentially, $ArchES$ measures the information leakage from an Arch-hint in UM system by examining the relation between the Arch-hints pattern (i.e., $distinguishability$, $consistency$) and the victim model internal architectures. 
As we analyze in Sec. \ref{unireveal}, the far page fault latency is closely associated with the OFM size of almost all layers, the migration latency and size can reveal the Filter size of a layer (i.e., Conv/FC). 
As the three primary Arch-hints in UM provide an ever explored attack surface for adversary, we will show below that they exhibit sufficient information for UMProbe to extract the model architecture. Then, the common Arch-hints of L2 read and write transaction can further enhance UMProbe performance by providing additional information to identify such blurring layers as BN and ACT layer. 

\begin{table}[t]
\centering
\small
\caption{Samples using different Arch-hints.}
\vspace{-6pt}
\begin{tabular}{c|c|c|c|c|c}
\hline
\tabincell{l}{Sample} &\tabincell{l}{$s_{1}$} &\tabincell{l}{$s_{2}$} &\tabincell{l}{$s_{3}$} &\tabincell{l}{$s_{4}$} &\tabincell{l}{$s_{5}$}\\\hline
\tabincell{c}{Arch-hints} &\tabincell{c}{PFLat}&\tabincell{c}{MigSize}&\tabincell{l}{PriArchs} &\tabincell{l}{ComArchs} &\tabincell{c}{AllArchs}\\\hline
\end{tabular}
\vspace{-2pt}
\label{archhintsample}
\end{table}

\begin{figure}[t]
\centering
            \includegraphics[width=10cm, height = 3.5cm, trim=120 0 0 0]{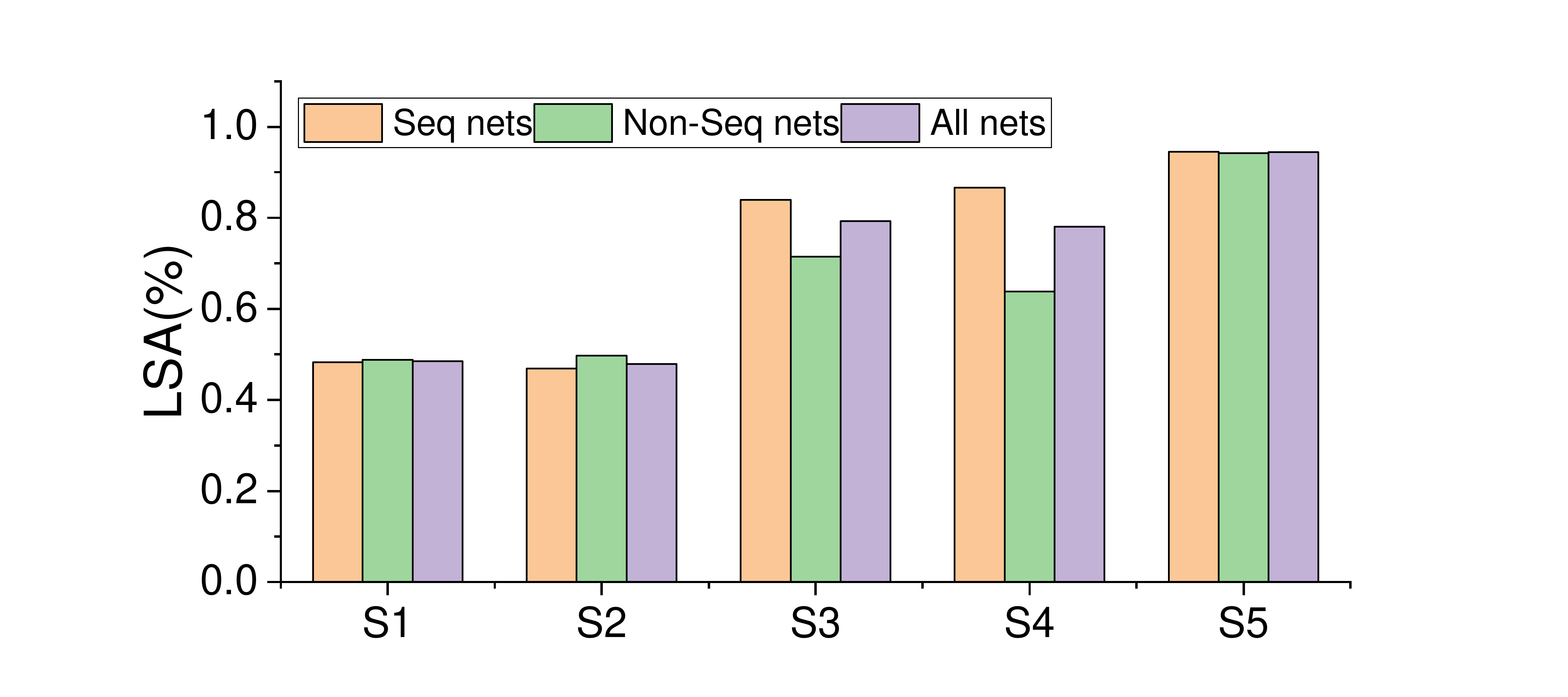}
            \vspace{-5mm}
            \caption{Avg LSA of UMProbe on different models.}\label{avglascom}
            \vspace{-16pt}
\end{figure}

\subsection{UMProbe Performance}\label{evalacc}
\vspace{-3pt}
\noindent\textbf{Metric:}
As UMProbe targets extracting the victim DNN model layer sequence (i.e., layer number, layer types and layer connection), we measure the performance of UMProbe's DNN extraction ability by quantifying the extracted layer sequence accuracy. We define the extracted \ul{\textit{Layer Sequence Accuracy (LSA)}} as follows,
\begin{equation}
\footnotesize
LSA = 1 - \frac{ED(L, L^{\ast})}{|L^{\ast}|}
\end{equation}\label{lac}
where $ED(L, L^{\ast}$) is the edited distance between extracted layer sequence $L$ and ground-truth layer sequence $L^{\ast}$ (i.e. the minimum number of insertions, substitutions, or deletions required to change L into $L^{\ast}$) \cite{abu2015exact}, while \textbf{$\left. ED(L, L^{\ast}) \middle/ |L^{\ast}| \right.$} indicates the extracted layer sequence error rate. $|L^{\ast}|$ is the length of $L^{\ast}$, thus, the larger $LSA$ is, the less the difference between $L$ and $L^{\ast}$, and the more accurate UMProbe extraction.

\begin{figure}[t]
\centering
            \includegraphics[width=10cm, height = 4.2cm, trim=90 0 0 0]{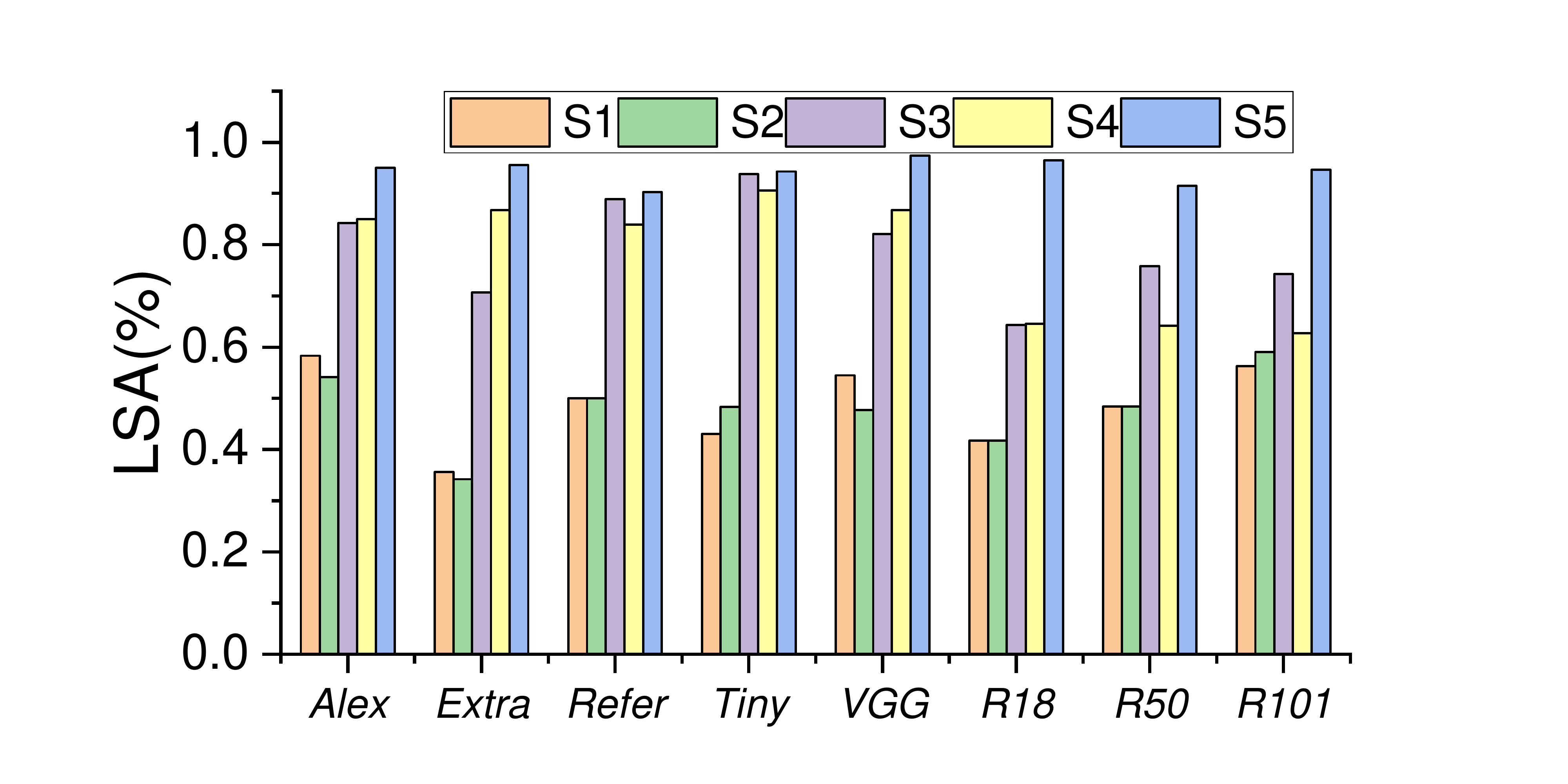}
            \vspace{-3mm}
            \caption{LSA of benchmarks using different Arch-hints.}\label{lascom}
            \vspace{-16pt}
\end{figure}

\noindent\textbf{Evaluation:}
As UMProbe works by leveraging different Arch-hints samples (see Table \ref{archhintsample}), different Arch-hints are able to reveal different DNN layer features and model characteristics, and make a great difference to UMProbe performance in terms of LSA. We measure UMProbe performance on DNN benchmarks, and further validate the importance and effectiveness of the Arch-hints in UM system. 

First, we calculate the the average LSA of UMProbe using different Arch-hints on three kinds of networks (i.e., Seq nets, Non-Seq nets and all nets), as shown in Fig.\ref{avglascom}. 
We observe that the LSA of UMProbe by using either $s_{1}$ or $s_{1}$ is around 50\%, indicating that UMProbe can effectively extract partial DNN layer sequence, though its performance is low.  
As we analyzed in Sec. \ref{unireveal}, all layers except BN/ACT can cause far page fault and exhibit PFLat, which is closely associated with the OFM size of the layer, while the MigSize closely correlates to the Filter size of a layer (i.e., the dominant Conv/FC). 
Thus, both Arch-hints can provide effective information for adversary to infer the DNN layer sequence, but the amount of information is limited. 

Then, we learned that, by using $s_{3}$ of PriArchs (i.e., PFLat, MigLat, MigSize), the UMProbe performance is obviously improved. 
As we analyzed above, PFLat is closely associated with a layer OFM size while MigLat/MigSize can reveal the Filter size of a layer, indicating the Arch-hints can provide complementary information about the DNN architecture. By using this three primary Arch-hints, UMProbe can effectively extract most layer sequence information.

Meanwhile, by using $s_{4}$ of ComArchs, the UMProbe performance is also improved, and is comparative to UMProbe using $s_{3}$. 
Regarding GPU memory hierarchy, the L2 cache is shared by all GPU SMs, while a kernel can be dispatched to multiple SMs and the kernel typically cannot bypass L2 cache to read/write data from DRAM. Thus, the L2 transaction provides relatively complete and highly distinguishable trace of data activities from different layer (i.e., input and output), as Table \ref{AvES} shows the high $ArchES$ of the Arch-hints. Thus, UMProbe performance using ComArchs is high as well. 
Based on the analysis above, we can say that as the new attack surface in UM system, the PriArchs provide sufficient information for UMProbe to effectively extract most of victim layer sequence, though UMProbe performance is not high enough.

Finally, by using $s_{5}$, the average LSA of UMProbe on Seq, Non-Seq, and all networks can reach around 95\%,
indicating that UMProbe can effectively extract almost all layer sequence.    
As we analyzed above, the PriArchs provide sufficient information for UMProbe to successfully extract layer sequence. Now, given that the ComArchs can provide additional information to further identify such blurring layers as BN/ACT, which hardly causes page fault and data migration, UMProbe performance can be further improved with the help of ComArchs.



Besides, we calculate UMProbe LSA on each DNN benchmark, as shown in Fig.\ref{lascom}.
We observe UMProbe performance on each DNN model that follows the same track of analysis above. Basically, PFLat reveals the OFM characteristics and MigSize reveals the Filter characteristics, either of them provides limited information for UMProbe ($\sim$ 50\% accuracy). Then, the three primary Arch-hints together can reveal a layer's features more completely, and UMProbe performance can be significantly improved. Especially, for the small and neat networks (e.g., Alexnet, Reference, Tiny and VGG), UMProbe performance is high ($\geq$ 80\%), indicating that the primary Arch-hints are able to reveal such model architecture thoroughly.  
Later, by using all Arch-hints, UMProbe can achieve very high performance on all models ($\geq$ 90\%).  

To summarize, we conclude that the new attack surface provided by the Arch-hints based on far page fault and data migration provides sufficiently effective clues for the adversary to extract most victim model architecture information in UM system and the extraction attack can achieve a high performance. Also, with Arch-hints providing additional information, the attack surface can be extended and the attack performance can be enhanced. In fact, such an attack surface has never been explored before and is worth attention.

\section{Related Work}
\noindent\textbf{Unified Memory:} 
GPU Unified Memory (UM) arises as it effectively eliminates the need for manual data migration, reducing programmer effort and enabling GPU memory oversubscription compared to the Copy-then-Execute system. However, the far fault handling and on demand migration can significantly impact the application performance, and many prior works focusing on performance optimization \cite{zheng2016towards, gandhi2014efficient, shin2018scheduling, hao2017supporting, kim2020batch, wang2020enabling, wang2020understanding, ganguly2019interplay} have been proposed. \cite{zheng2016towards} proposes a software page prefetcher to further utilize PCIe bus bandwidth and hide page migration overheads. \cite{kim2020batch} comprehensively characterizes the inefficiency of far fault handling under UM model and proposes batch-aware UM management. \cite{ganguly2019interplay} investigates the prefeching and eviction policies under UM model and proposes new locality-aware pre-eviction policies to reduce the performance overhead. However, this paper first explores the insecure communication pattern exposed by the far fault handling and on-demand migration under UM model and exploits this attack surface in UM system for stealing DNN models.   

\noindent\textbf{Model Extraction Attack:} 
The extraction attack mainly targets the ML models deployed in cloud with publicly accessible query inter-faces/APIs, and the adversary can duplicate the functionality of the model by frequently querying APIs \cite{oh2019towards, tramer2016stealing}. 
Then, some works consider utilizing side-channel information to benefit the attacks \cite{hong2018security, hong20200wn, yan2020cache, hua2018reverse}, such as cache-side channel.   
Recently, with ML models increasingly deployed in edge/local devices \cite{li2018learning, yazici2018edge, verhelst2020machine}, the adversary utilizes physical or local side-channels to obtain architecture-level information leakage to accurately extract the model architecture.
\cite{naghibijouybari2018rendered} utilizes hardware counters
to predict the NN neuron number. \cite{wei2020leaky} monitors the CUPTI events in GPU platforms to infer different DNN layer operations.
\cite{hua2018reverse} observe the memory access patterns to search for the possible DNN structures in FPGAs. 
\cite{hu2020deepsniffer} collects the kernel latency, DRAM read and write volume, etc., to extract the DNN model architectures. 
However, none of them explores the meanings and patterns behind architecture information or proposes new architecture hints and attack surface for extraction attack in UM system.

\noindent\textbf{Mitigation Countermeasures:}
As the new attack surface relies on insecure communication pattern between GPU and CPU on PCIe bus, one potential defense approach is to obfuscate the communication pattern on PCIe bus. 
As GPU runtime process the far page fault first and then migrate data on demand, the runtime can dynamically obfuscate the requests, like postpone or even reorder some far fault requests. Also, the runtime/system can support transmitting dummy data to cover the real traffic,
for example, the migrated data can be split/padded into a fixed size and be sent at fixed rate \cite{hunt2019isolation}. This way, the PCIe transmission and leaky communication pattern in UM system can obfuscated and interfered.
However, such approaches will unavoidably incur significant PCIe bandwidth overhead and performance degradation. 

Besides, GPU trust execution environment (TEE) can be considered to mitigate or eliminate co-location side channel \cite{hunt2020telekine, volos2018graviton, 2020towards}. These TEE disallows different tenants to share the underlying hardware or execute concurrently, which can prevent the adversary to observe the victim activities through the performance counter, etc. Similarly, this method can negatively impact the GPU performance and is non-trivial to be deployed in practice. 

\vspace{-1em}
\section{Conclusion}
Emerging extraction attack can leverage architecture-level events (i.e., Arch-hints) in hardware platforms to extract DNN model layer information accurately. 
In this paper, we uncover the root cause of such Arch-hints and summarize the principles to identify them. We then apply these principles to emerging Unified Memory (UM) management system, identify three new Arch-hints, and develop a new extraction attack, UMProbe. We also create the first DNN benchmark suite in UM and utilize the benchmark suite to evaluate UMProbe. Evaluation shows UMProbe can extract the layer sequence with an accuracy of 95\% for almost all victim test models, calling for more attention to the DNN security in UM system.



\end{document}